\documentclass[a4paper,12pt]{article}
\usepackage{amsmath}
\usepackage{amssymb}
\usepackage{graphicx}
\usepackage{longtable}
\usepackage{multirow}
\usepackage{array}
\usepackage[hidelinks]{hyperref} 
\usepackage[affil-it]{authblk}
\usepackage{booktabs}
\usepackage[flushmargin, hang]{footmisc}
\usepackage[a4paper, margin=2cm]{geometry}
\usepackage{natbib}
\setcitestyle{authoryear,open={(},close={)}}
\usepackage{tikz}
\usetikzlibrary{shapes.geometric, arrows.meta, positioning, calc}
\usepackage{pdflscape}

\begingroup
\catcode`\,=\active
\gdef,{\mathchar"613B\allowbreak}
\endgroup
\DeclareSymbolFont{sfletters}{OT1}{cmss}{m}{n}
\newcommand{\fml}[1]{{\scriptsize\binoppenalty=0\relpenalty=0\mathcode`\,="8000 $\fam\symsfletters #1$}}

\title{Cross-Market Alpha:\\Testing Short-Term Trading Factors in the U.S.\ Market via Double-Selection LASSO}

\author{
    Jin Du\textsuperscript{*,$\dagger$},
    Alexander Walter\textsuperscript{*,\,$\ddagger$},
    Maxim Ulrich\textsuperscript{$\S$}
}

\date{
    \vspace{1em}
    \begin{minipage}{\textwidth}
        \centering
        \small
        \begin{tabular}{r@{\hspace{0.5em}}l}
            \textsuperscript{*} Equal contribution \\
            \textsuperscript{$\dagger$} \texttt{jin.du@partner.kit.edu} \\
            \textsuperscript{$\ddagger$} \texttt{alexander.walter@partner.kit.edu} (Corresponding author) \\
            \textsuperscript{$\S$} \texttt{maxim.ulrich@kit.edu} \\
        \end{tabular}
    \end{minipage}
    \\[2em]
    \today
}

\begin{document}

\maketitle

\begin{abstract}
We test whether 168 short-horizon price-volume signals from the Alpha191 library, originally developed for China's retail-dominated A-share market, contain pricing information for S\&P 500 stocks from 2002 to 2022 beyond 153 established U.S.\ factors. Using the double-selection LASSO of \citet{feng2020taming}, 17 signals receive significant stochastic discount factor (SDF) loadings in the baseline test-asset design. Their robustness is uneven. Only three signals (a multi-horizon moving-average ratio, a directional-pressure ratio, and a price-gap correlation) remain significant with a finer test-asset grid and under Elastic Net and principal-component control selection; six more pass most checks, and the remaining eight depend on the specification. Robust signals are concentrated in volume-price interaction and short-term mean reversion, whereas volatility-based signals are fragile.
\end{abstract}

\vspace{1\baselineskip}

\vspace{1\baselineskip}
\noindent\textbf{Key Findings}
\begin{itemize}
    \item Of 168 Chinese A-share trading signals, 17 add significant pricing information for S\&P 500 stocks beyond 153 established U.S.\ factors in a double-selection LASSO framework.
    \item Robustness is concentrated: three signals survive every alternative specification, six more survive most, and eight depend on the specification.
    \item Volume-price interaction and short-term mean-reversion signals transfer best across markets; volatility-based signals do not survive alternative test assets or control selection.
\end{itemize}

\newpage

\section{Introduction}

A large literature has produced hundreds of candidate return predictors, raising well-known concerns about data mining and multiple testing \citep{harvey2016and}. Most of this factor zoo \citep{jensen2023replication} consists of characteristics built from slow-moving accounting data or from monthly and longer return histories. Short-horizon trading signals built from daily prices and volumes are comparatively under-represented. In this article, we ask whether such signals, taken from a library developed for a very different market, contain pricing information for large U.S.\ stocks that the established factor zoo does not already capture.

Our source of signals is the Alpha191 library published by Guotai Junan Securities for the Chinese A-share market. That market is dominated by retail trading. According to annual reports of the China Securities Depository and Clearing Corporation, retail investors account for roughly 60\% of trading volume. The library catalogs 191 price-volume and order-flow formulas designed for this high-turnover environment. We apply them to S\&P 500 constituents from 2002 to 2022, a segment that is institutionally dominated, highly liquid, and heavily researched. If a signal carries information in this setting after controlling for 153 characteristics from \citet{jensen2023replication}, it is unlikely to reflect small-stock illiquidity or a subset of well-known anomalies. To handle the high-dimensional and collinear control set, we use the double-selection (DS) LASSO of \citet{belloni2014inference} and \citet{feng2020taming}.

We obtain three main findings. First, 17 of 168 signals receive significant SDF loadings in the baseline $3\times2$ design. Six of them have $t$-statistics above 3, the threshold proposed by \citet{harvey2016and} for new factors, and two survive a Bonferroni correction for 168 tests. Second, robustness is concentrated in a small core. Only three signals remain significant, with the same sign, in a finer $5\times5$ test-asset design and under Elastic Net and principal-component control selection: the Multi-Period Mean Reversion Ratio (046), the Downward Directional Pressure Ratio (049), and the Rank of Delayed Price-Gap Correlation (184). Six further signals pass two of the three checks without a significant sign reversal, and the remaining eight depend on the specification. Third, survival differs systematically by signal type. Volume-price interaction and short-term mean-reversion signals are over-represented among robust signals, while volatility-based signals appear in the baseline but disappear under alternative specifications.

The article makes three contributions. First, it provides a cross-market test of a complete, publicly documented trading-signal library against a comprehensive U.S.\ control set, using an estimator designed for high-dimensional factor selection. Second, it offers a transparent robustness scorecard that separates a small set of stable signals from specification-dependent ones. This scorecard is a practical tool for anyone screening large signal libraries under multiple-testing concerns. Third, it shows which categories of short-horizon signals transfer from China to U.S.\ large caps, and it relates this pattern to how well each category is already covered by the U.S.\ factor zoo.

The remainder of the article is organized as follows. Section~\ref{sec:data} describes the data, signal construction, and portfolio timing. Section~\ref{sec:method} presents the DS-LASSO methodology and the interpretation of its coefficients. Section~\ref{sec:results} reports the baseline results, the robustness scorecard, and the category pattern. Section~\ref{sec:discussion} discusses economic interpretations, Section~\ref{sec:limitations} states the limitations of the analysis, and Section~\ref{sec:conclusion} concludes. An online appendix contains the full robustness tables, a horizon diagnostic, and the definitions of all signals.

\newpage

\section{Data and Signal Construction}\label{sec:data}

\subsection{Data}

\begin{figure}[ht!]
    \centering
    \includegraphics[width=0.55\textwidth]{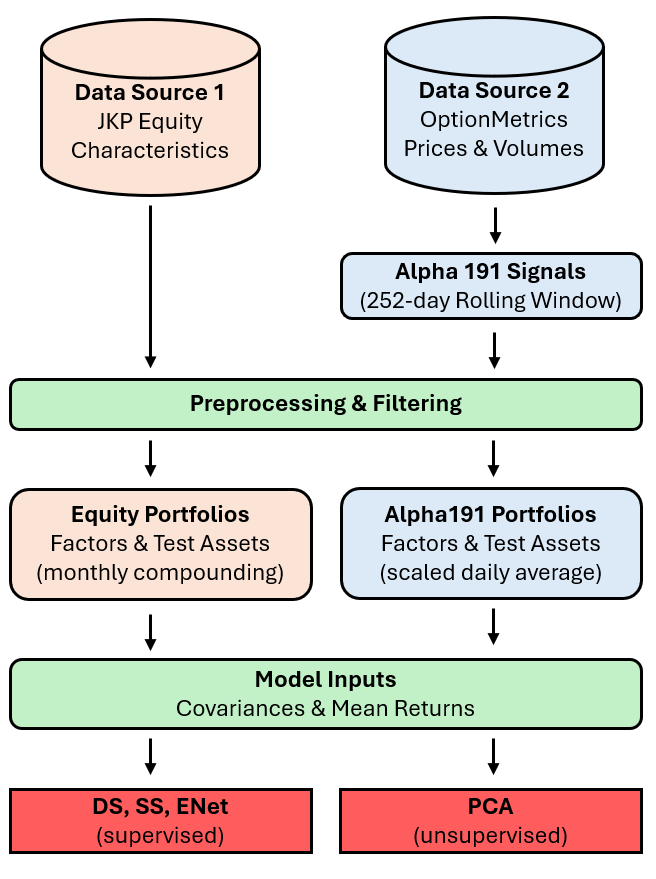}
    \caption{Empirical workflow}
    \label{fig:empirical_workflow}
\end{figure}

Figure~\ref{fig:empirical_workflow} summarizes the empirical workflow. The control universe comes from the global factor database of \citet{jensen2023replication}. For the U.S., this database contains 4,135,225 firm-month observations with 444 variables. We retain CRSP common shares (share codes 10 and 11) listed on NYSE, AMEX, or NASDAQ with a price above \$5, and we use the 153 characteristics that define the \citet{jensen2023replication} factors.

The Alpha191 signals are computed from daily open, high, low, and close prices and trading volumes of S\&P 500 constituents in OptionMetrics IvyDB, with 4,917,327 stock-day observations from 2002 to 2022. Daily index membership is used to determine the constituents on each date. Each signal is computed from a rolling 252-trading-day window, and stocks without a complete window are excluded. We classify the 191 formulas into six thematic groups: Volume \& Flow, Mean Reversion, Trend \& Momentum, Volatility \& Risk, Liquidity \& VWAP, and Price Action. Many formulas could arguably belong to more than one group, so the classification is a heuristic for interpretation rather than a structural decomposition. We exclude 23 signals whose time series contain excessive missing values or are numerically unstable in the rolling computation, which leaves 168 signals for the analysis. Online Appendix Table~A4 lists all formulas, their categories, and the exclusions.

\subsection{Portfolio Construction and Timing}

All factor portfolios are rebalanced monthly. At the close of the last trading day of month $t$, we compute each signal for all stocks that are S\&P 500 constituents on that day. Stocks are sorted by the signal, and the factor is long the top decile and short the bottom decile, with value weights based on market capitalization on the formation date. The portfolio is held unchanged throughout month $t+1$, and we record its daily long-short returns. The monthly factor return is the average daily long-short return in month $t+1$ multiplied by 21. For example, holdings formed at the close of 29 January 2010 earn the daily returns from 1 February to 26 February 2010 and are replaced at the close of 26 February.

The \citet{jensen2023replication} characteristics are available only at monthly frequency. The corresponding control factors are therefore built as value-weighted high-minus-low decile portfolios from lagged characteristics and market capitalizations, using monthly returns directly. The DS-LASSO operates on covariances between test-asset returns and factor returns, so control factors, Alpha191 factors, and test assets must be observed at a common frequency. This data constraint, rather than a choice about signal horizon, determines the monthly frequency of the analysis.

Our monthly Alpha191 factor return is the scaled arithmetic average of daily returns rather than the compounded monthly return. The two differ only through the compounding (volatility-drag) term. A paired $t$-test cannot reject equality of the two series for 134 of the 168 signals ($p>0.05$), including 13 of the 17 signals that are significant in our baseline results. The choice of aggregation therefore has little influence on the main results.

\subsection{Test Assets}

Test assets are formed by independent bivariate sorts on size and on each signal or characteristic. They are value-weighted, rebalanced monthly on the same dates as the factors, and drawn from the set of stocks that are S\&P 500 constituents at any point in the sample. For characteristics that take discrete values with many ties, such as firm age in years, we use overlapping bins so that each portfolio contains enough stocks. The baseline $3\times2$ design produces 1,008 Alpha191 portfolios ($168\times6$) and 918 portfolios based on the \citet{jensen2023replication} characteristics ($153\times6$), for a total of 1,926 test assets. The finer $5\times5$ design produces 4,200 and 3,825 portfolios, for a total of 8,025 test assets. Monthly value-weighted portfolio returns are the dependent variables in all regressions.

\newpage

\section{Methodology}\label{sec:method}

Evaluating a new factor against a large set of existing ones raises two linked problems. The control set is high-dimensional and highly collinear, and omitting a control that is correlated with the tested factor biases its estimated contribution. A single LASSO selection step can drop controls that matter little for average returns but are strongly correlated with the tested factor \citep{belloni2014inference, fastrich2015constructing}. The DS-LASSO of \citet{feng2020taming} addresses this problem with two selection steps followed by a post-selection regression (Figure~\ref{fig:methodology}).

\begin{figure}[ht!]
\centering
\begin{tikzpicture}[
    box/.style={draw, rounded corners=4pt, thick, fill=blue!6, align=center, text width=6.3cm, minimum height=1.9cm, font=\small},
    wide/.style={draw, rounded corners=4pt, thick, fill=blue!6, align=center, text width=13.9cm, font=\small},
    arr/.style={-{Stealth[length=2.5mm]}, thick}
]
\node[box] (s1) {\textbf{Step 1 ($I_1$): return drivers}\\ controls $h$ whose covariances $\widehat{\mathrm{Cov}}(r,h)$ explain average returns $\bar r$};
\node[box, right=0.8cm of s1] (s2) {\textbf{Step 2 ($I_2$): confounders}\\ controls $h$ whose covariances explain $\widehat{\mathrm{Cov}}(r,g_j)$ for each tested signal $g_j$};
\node[wide, below=0.9cm of $(s1.south)!0.5!(s2.south)$, minimum height=1.0cm] (u) {\textbf{Union of selected controls} $I_1 \cup I_2$};
\node[wide, below=0.9cm of u, minimum height=1.6cm] (s3) {\textbf{Step 3: post-selection OLS}\\ $\bar r \sim \text{const} + \widehat{\mathrm{Cov}}(r, h_{I_1\cup I_2})\,\lambda_h + \widehat{\mathrm{Cov}}(r, g)\,\lambda_g$ \quad $\Rightarrow$ SDF loadings $\hat\lambda_g$ of the tested signals};
\draw[arr] (s1.south) -- (s1.south |- u.north);
\draw[arr] (s2.south) -- (s2.south |- u.north);
\draw[arr] (u) -- (s3);
\end{tikzpicture}
\caption{Double-selection LASSO: overview}
\label{fig:methodology}
\end{figure}

\paragraph{Step 1: Selection of return drivers.} A cross-sectional LASSO regression of average test-asset returns on their covariances with the control factors identifies the controls that explain the cross-section of expected returns:
\begin{equation}
\min_{\gamma,\lambda}\left\{n^{-1}\left\|\bar{r}-\iota_n\gamma-\widehat{\mathrm{Cov}}\left(\boldsymbol{r}_t,\boldsymbol{h}_t\right)\lambda\right\|^2+\tau_0 n^{-1}\|\lambda\|_1\right\}.
\label{eq:1}
\end{equation}
Here, $\bar{r}$ is the vector of time-series mean returns of the $n$ test assets, $\iota_n$ is a vector of ones, $\gamma$ is the zero-beta rate, $\widehat{\mathrm{Cov}}(\boldsymbol{r}_t,\boldsymbol{h}_t)$ is the $n\times p$ matrix of sample covariances between test-asset returns and the $p=153$ control factors, and $\tau_0$ is the penalty parameter. Controls with non-zero coefficients form the set $I_1$.

\paragraph{Step 2: Selection of confounders.} For each tested signal $j=1,\dots,d$, a second LASSO regression of its covariances with the test assets on the control covariances identifies the controls that are correlated with that signal:
\begin{equation}
\min_{\xi_j,\chi_j}\left\{n^{-1}\left\|\widehat{\mathrm{Cov}}\left(\boldsymbol{r}_t,g_{t,j}\right)-\iota_n\xi_j-\widehat{\mathrm{Cov}}\left(\boldsymbol{r}_t,\boldsymbol{h}_t\right)\chi_j\right\|^2+\tau_{1j} n^{-1}\|\chi_j\|_1\right\}.
\label{eq:2}
\end{equation}
Step 2 is run over the full control universe, not only over $I_1$. Controls with non-zero coefficients in any of the $d$ regressions form the set $I_2$. This step retains controls that may be weak drivers of average returns but would bias the estimate for a signal if omitted.

\paragraph{Step 3: Post-selection inference.} Finally, average returns are regressed by OLS on the covariances with the selected controls and with the tested signals:
\begin{equation}
\begin{split}
\left(\hat{\gamma}_0,\hat{\lambda}_h,\hat{\lambda}_g\right)=\arg\min_{\gamma_0,\lambda_h,\lambda_g}\Big\{&\left\|\bar{r}-\iota_n\gamma_0-\widehat{\mathrm{Cov}}\left(\boldsymbol{r}_t,\boldsymbol{h}_t\right)\lambda_h-\widehat{\mathrm{Cov}}\left(\boldsymbol{r}_t,\boldsymbol{g}_{t}\right)\lambda_g\right\|^2 :\\
&\lambda_{h,i}=0\ \ \forall\, i\notin I_1\cup I_2\Big\}.
\end{split}
\label{eq:3}
\end{equation}
We report heteroscedasticity-robust (HC3) $t$-statistics.\footnote{Implementation details: the intercepts in Equations~\eqref{eq:1} and \eqref{eq:2} are unpenalized, so selection is driven by the cross-sectional covariance structure rather than by the level of average returns. The penalties $\tau_0$ and $\tau_{1j}$ are chosen by 10-fold cross-validation (scikit-learn \texttt{LassoCV}, 200 candidate penalties, \texttt{eps} $=0.05$) using the one-standard-error rule, i.e., the largest penalty whose cross-validated mean squared error is within one standard error of the minimum. This rule yields sparser control sets than the MSE-minimizing penalty. Computations use Python 3.12.0 and scikit-learn 1.7.1.} Unlike \citet{feng2020taming}, who evaluate one new factor at a time, we include all 168 Alpha191 signals jointly in Step 3, together with the union of all Step-2 selections. Each $\hat\lambda_{g,j}$ is therefore conditional on the selected controls and on the other 167 signals.

\paragraph{Interpretation of $\lambda$.} Following \citet{feng2020taming}, $\lambda_g$ is a loading of the SDF on the factor, not a risk premium. It measures whether the factor helps price the test assets, given the other factors in the model. A factor whose own long-short portfolio earns close to zero on average can have a significant SDF loading if its covariance with the test assets explains average returns that the other factors leave unexplained, for example because it offsets exposures of correlated factors. Conversely, a factor with a large average return can have a zero loading if other factors already span it. All factors are standardized to unit variance, so $\lambda$ is measured in basis points of monthly expected return per unit of covariance with a one-standard-deviation factor, which makes magnitudes comparable across signals. In our joint specification, $\lambda$ is best read as the marginal contribution of a signal to pricing the cross-section, conditional on everything else in the model.

\paragraph{Why DS-LASSO.} Alternative high-dimensional methods trade off interpretability and inference differently. Principal components summarize the controls in latent factors, which removes collinearity but no longer identifies individual controls. Elastic Net handles correlated regressors through joint shrinkage but has no second selection step targeted at confounders of the tested factors \citep{lettau2020factors, gu2020empirical}. We therefore use DS-LASSO as the baseline and report Elastic Net and principal-component versions as robustness checks.

\newpage

\section{Results}\label{sec:results}

\subsection{Baseline Results}

Table~\ref{tab:3x2_DS_SS_results} reports the 17 Alpha191 signals with SDF loadings that are significant at the 5\% level in the baseline $3\times2$ design. For comparison, column (2) shows single-selection (SS) estimates, which omit the second selection step.

\setlength{\tabcolsep}{4pt}
\renewcommand{\arraystretch}{1.2}

\begingroup\small
\begingroup\small
\begin{longtable}{
    >{\raggedright\arraybackslash}p{0.43\textwidth}
    >{\centering\arraybackslash}p{0.12\textwidth}
    >{\centering\arraybackslash}p{0.12\textwidth}
    >{\centering\arraybackslash}p{0.12\textwidth}
    >{\centering\arraybackslash}p{0.12\textwidth}
}
\caption{SDF loadings of Alpha191 signals, $3\times2$ test assets \label{tab:3x2_DS_SS_results}} \\
\toprule
~ & \multicolumn{2}{c}{(1) DS-LASSO} & \multicolumn{2}{c}{(2) SS-LASSO} \\
\cmidrule(lr){2-3} \cmidrule(lr){4-5}
Signal & $\lambda$ (bp) & $t$ & $\lambda$ (bp) & $t$ \\
\midrule
\endfirsthead
\caption*{Table \thetable{} -- continued} \\
\toprule
Signal & $\lambda$ (bp) & $t$ & $\lambda$ (bp) & $t$ \\
\midrule
\endhead
\midrule
\multicolumn{5}{r}{Continued on next page\ldots} \\
\endfoot
\bottomrule
\multicolumn{5}{p{0.93\textwidth}}{\footnotesize \textit{Note:} $\lambda$ is the SDF loading on the covariance with the standardized factor, in basis points per month. HC3 $t$-statistics. $^{*}$ and $^{**}$ indicate significance at the 5\% and 1\% levels, respectively.} \\
\endlastfoot
Multi-Period Mean Reversion Ratio (046) & 79 & $3.68^{**}$ & -19 & $-0.12$ \\
20-Day Cumulative On-Balance Volume (084) & 51 & $3.68^{**}$ & -1052 & $-3.14^{**}$ \\
Inverse Rank of Nested Decayed Price-Volume Correlations (073) & 40 & $3.41^{**}$ & -50 & $-1.02$ \\
Price-Volume vs. Low-Volume Correlation Rank (123) & 42 & $3.39^{**}$ & 217 & $3.66^{**}$ \\
Downward Directional Pressure Ratio (049) & 25 & $3.12^{**}$ & -90 & $-2.42^{*}$ \\
24-Day Percentage Deviation from Mean (071) & 58 & $3.06^{**}$ & -300 & $-1.36$ \\
Rank of Delayed Price-Gap Correlation (184) & 41 & $2.86^{**}$ & 18 & $0.22$ \\
Volume MACD Histogram (155) & 40 & $2.82^{**}$ & -150 & $-3.38^{**}$ \\
Inverse Rank of Intraday Volatility and Corr (054) & 44 & $2.67^{**}$ & 800 & $2.99^{**}$ \\
Benchmark-Relative Excess Return Skewness (181) & 38 & $2.57^{*}$ & 243 & $4.51^{**}$ \\
12-Day Average True Range (161) & 34 & $2.57^{*}$ & 4286 & $3.89^{**}$ \\
Log Gain-to-Loss Variability Ratio (190) & 36 & $2.49^{*}$ & -593 & $-3.64^{**}$ \\
Rank of Decay-Adjusted Momentum-VWAP Divergence (039) & 39 & $2.49^{*}$ & 478 & $3.38^{**}$ \\
Overnight Gap Return (015) & 41 & $2.46^{*}$ & -270 & $-0.75$ \\
6-Day Relative Strength Index (063) & 45 & $2.38^{*}$ & -212 & $-2.07^{*}$ \\
6-Day Negative Correlation of Volume Growth and Return (001) & 37 & $2.31^{*}$ & 558 & $4.69^{**}$ \\
10-Day Price Acceleration vs. Directional Change (086) & 32 & $2.30^{*}$ & 158 & $3.11^{**}$ \\
\end{longtable}

\newpage

All 17 significant DS loadings are positive, with magnitudes between 25 and 79 basis points. Six signals have $t$-statistics above 3, the hurdle suggested by \citet{harvey2016and} to account for multiple testing in the factor literature. A Bonferroni correction for 168 tests at the 5\% level requires $|t|>3.62$, which only the Multi-Period Mean Reversion Ratio (046) and the 20-Day Cumulative On-Balance Volume (084) meet. The remaining signals should be read as candidates whose evidence depends on robustness across specifications, which we examine in Section~\ref{sec:robustness}.

The SS estimates differ from the DS estimates in sign or significance for most signals, and several SS magnitudes are implausibly large. Because SS omits controls that are correlated with the tested signals, these differences show that inference on individual signals is sensitive to control selection. They do not by themselves prove that the DS estimates are correct, which is a further reason to evaluate the signals across several specifications.

Several baseline signals have own long-short returns that are close to zero or even significantly negative, as Table~\ref{tab:scorecard} shows. This is consistent with the SDF-loading interpretation in Section~\ref{sec:method}. For example, the Multi-Period Mean Reversion Ratio (046) earns an average long-short return that is statistically indistinguishable from zero ($t=0.21$), yet it has the largest SDF loading. The Inverse Rank of Intraday Volatility (054) and the 12-Day Average True Range (161) have significantly negative average returns but positive loadings. These signals matter for pricing because of how their returns covary with the test assets, given the other factors, and not because their portfolios earn a premium on their own.

\subsection{Robustness and the Scorecard}\label{sec:robustness}

We check the baseline results in two ways. First, we replace the $3\times2$ test assets with the finer $5\times5$ design, which quadruples the number of test assets and increases cross-sectional power. Second, we keep the $3\times2$ design but select the controls with cross-validated Elastic Net or replace them with the principal components that explain 90\% of the variance of the control covariances. In both cases, all Alpha191 signals are included in the final OLS regression. The full results are reported in Online Appendix Tables~A1 and A2.

The finer design yields 30 significant signals, but only 8 of the 17 baseline signals are among them. Twenty-two signals are significant only in the $5\times5$ design. Elastic Net and PCA each confirm 9 of the 17 baseline signals with the same sign, and only 5 are confirmed by both. Three signals reverse sign significantly: 073 and 161 under Elastic Net and 063 under PCA. The Elastic Net magnitudes are often an order of magnitude larger than the DS estimates, which reflects the different conditioning set rather than a different economic effect. Taken together, the set of significant signals clearly depends on the specification, and the baseline list of 17 should not be read as 17 robust factors.

Table~\ref{tab:scorecard} organizes this evidence into a robustness scorecard. For each baseline signal, it reports the DS $t$-statistic in the $3\times2$ and $5\times5$ designs, the Elastic Net and PCA $t$-statistics, and the $t$-statistic of the signal's own average long-short return. We classify signals into three tiers:
\begin{itemize}
    \item \textit{Robust core}: significant with the same sign in all three alternative specifications.
    \item \textit{Second tier}: significant with the same sign in two of three, with no significant sign reversal.
    \item \textit{Fragile}: all other signals.
\end{itemize}

\newpage

\begin{table}[ht!]
\centering
\caption{Robustness scorecard for the 17 baseline signals}
\label{tab:scorecard}
\footnotesize
\setlength{\tabcolsep}{4pt}
\renewcommand{\arraystretch}{1.15}
\resizebox{\textwidth}{!}{%
\begin{tabular}{l l c c c c c c}
\toprule
 & & \multicolumn{2}{c}{DS-LASSO} & ENet & PCA & Own & Checks \\
\cmidrule(lr){3-4}
Signal & Category & $3\times2$ & $5\times5$ & $3\times2$ & $3\times2$ & return & passed \\
\midrule
\multicolumn{8}{l}{\textit{Robust core}} \\
Multi-Period Mean Reversion Ratio (046) & Mean Reversion & 3.68$^{\ddagger}$ & \textbf{4.33} & \textbf{4.34} & \textbf{5.29} & 0.21 & 3/3 \\
Downward Directional Pressure Ratio (049) & Trend \& Mom. & 3.12$^{\dagger}$ & \textbf{3.41} & \textbf{4.28} & \textbf{2.10} & 0.22 & 3/3 \\
Rank of Delayed Price-Gap Corr.\ (184) & Trend \& Mom. & 2.86 & \textbf{6.78} & \textbf{3.46} & \textbf{5.06} & 1.55 & 3/3 \\
\addlinespace
\multicolumn{8}{l}{\textit{Second tier}} \\
Price-Vol.\ vs.\ Low-Vol.\ Corr.\ Rank (123) & Volume \& Flow & 3.39$^{\dagger}$ & \textbf{5.43} & 1.15 & \textbf{3.52} & $-0.18$ & 2/3 \\
24-Day Pct.\ Deviation from Mean (071) & Mean Reversion & 3.06$^{\dagger}$ & n.s. & \textbf{4.00} & \textbf{2.76} & 0.30 & 2/3 \\
Volume MACD Histogram (155) & Volume \& Flow & 2.82 & \textbf{3.67} & \textbf{2.02} & $-0.05$ & 0.64 & 2/3 \\
Inv.\ Rank of Intraday Vol.\ and Corr.\ (054) & Volatility \& Risk & 2.67 & n.s. & \textbf{3.52} & \textbf{2.79} & $-2.78$ & 2/3 \\
Overnight Gap Return (015) & Trend \& Mom. & 2.46 & \textbf{5.17} & \textbf{3.94} & 1.02 & 2.71 & 2/3 \\
6-Day Neg.\ Corr.\ of Vol.\ Growth and Ret.\ (001) & Volume \& Flow & 2.31 & \textbf{3.41} & 0.74 & \textbf{2.33} & 0.37 & 2/3 \\
\addlinespace
\multicolumn{8}{l}{\textit{Fragile}} \\
20-Day Cumulative On-Balance Volume (084) & Volume \& Flow & 3.68$^{\ddagger}$ & n.s. & \textbf{4.43} & 0.55 & 0.61 & 1/3 \\
Inv.\ Rank of Nested Decayed P-V Corr.\ (073) & Volume \& Flow & 3.41$^{\dagger}$ & \textbf{4.29} & $-3.05^{\S}$ & \textbf{2.10} & 0.76 & 2/3$^{\S}$ \\
Benchmark-Rel.\ Excess Return Skewness (181) & Volatility \& Risk & 2.57 & n.s. & $-1.38$ & \textbf{4.94} & 1.82 & 1/3 \\
6-Day Relative Strength Index (063) & Trend \& Mom. & 2.38 & n.s. & \textbf{2.53} & $-2.94^{\S}$ & $-0.71$ & 1/3$^{\S}$ \\
12-Day Average True Range (161) & Volatility \& Risk & 2.57 & n.s. & $-3.94^{\S}$ & 1.12 & $-2.91$ & 0/3$^{\S}$ \\
Log Gain-to-Loss Variability Ratio (190) & Volatility \& Risk & 2.49 & n.s. & $-0.38$ & 1.44 & 0.14 & 0/3 \\
Rank of Decay-Adj.\ Mom.-VWAP Div.\ (039) & Liquidity \& VWAP & 2.49 & n.s. & 0.02 & 1.29 & $-1.08$ & 0/3 \\
10-Day Price Accel.\ vs.\ Dir.\ Change (086) & Trend \& Mom. & 2.30 & n.s. & 1.30 & 0.47 & 2.51 & 0/3 \\
\bottomrule
\end{tabular}}

\vspace{0.4em}
\begin{minipage}{0.97\textwidth}
\footnotesize \textit{Note:} Entries are $t$-statistics. Bold entries are significant at the 5\% level with the same sign as the baseline estimate. ``n.s.'': not significant at the 5\% level in the $5\times5$ design. ``Own return'': $t$-statistic of the signal's scaled monthly long-short return (Online Appendix Table~A3). ``Checks passed'': number of alternative specifications ($5\times5$, ENet, PCA) in which the signal is significant with the same sign. $^{\dagger}$: $|t|>3$ \citep{harvey2016and}. $^{\ddagger}$: $|t|>3.62$, Bonferroni-significant for 168 tests at the 5\% level. $^{\S}$: significant sign reversal, which places the signal in the fragile tier regardless of the number of checks passed.
\end{minipage}
\end{table}

Three results stand out. First, the robust core (046, 049, 184) consists of signals that pass every check, although only 046 also clears the Bonferroni threshold in the baseline. Second, a high baseline $t$-statistic is a poor predictor of robustness. The On-Balance Volume signal (084) is Bonferroni-significant in the baseline, but it is confirmed only by Elastic Net. Signal 073 has $t=3.41$, but it reverses sign significantly under Elastic Net. Third, the own-return column is unrelated to robustness. The three core signals have insignificant own average returns, which illustrates again that the DS loadings measure pricing contributions rather than stand-alone profitability.

\subsection{Which Signals Transfer?}

Table~\ref{tab:categories} counts signals by category at each stage of the analysis. It compares each category's share of the library with its share among the baseline signals, the robust core and second tier, and the signals significant in the $5\times5$ design.

\begin{table}[ht!]
\centering
\caption{Signals by category}
\label{tab:categories}
\small
\begin{tabular}{l c c c c}
\toprule
Category & Library & Significant & Core + & Significant \\
 & & $3\times2$ DS & second tier & $5\times5$ DS \\
\midrule
Trend \& Momentum  & 56 & 5 & 3 & 9 \\
Volume \& Flow     & 41 & 5 & 3 & 9 \\
Mean Reversion     & 26 & 2 & 2 & 7 \\
Volatility \& Risk & 24 & 4 & 1 & 1 \\
Liquidity \& VWAP  & 17 & 1 & 0 & 4 \\
Price Action       &  4 & 0 & 0 & 0 \\
\midrule
Total              & 168 & 17 & 9 & 30 \\
\bottomrule
\end{tabular}

\vspace{0.4em}
\begin{minipage}{0.8\textwidth}
\footnotesize \textit{Note:} ``Library'' counts the 168 signals used in the analysis. ``Core + second tier'' refers to the classification in Table~\ref{tab:scorecard}.
\end{minipage}
\end{table}

Two patterns emerge. Mean-reversion signals transfer best. They make up 15\% of the library but 22\% of the robust set and 23\% of the $5\times5$ survivors, and both baseline mean-reversion signals are robust. Volume \& Flow signals are also over-represented in the robust set. Volatility \& Risk signals show the opposite pattern. They are over-represented among the baseline survivors, with 4 of 17, but only one of them is robust, and only one volatility signal of 24 is significant in the $5\times5$ design. Trend \& Momentum signals roughly match their share of the library.

A plausible explanation for this pattern is how well each category is already covered by the U.S.\ control set. The \citet{jensen2023replication} characteristics include many volatility and risk measures, such as idiosyncratic volatility, market beta, maximum daily return, and return skewness. A volatility-based trading signal is therefore close to several controls, and its estimated contribution depends on which of them are selected, which is exactly the instability we observe. By contrast, the control set measures liquidity through levels, such as turnover, dollar volume, and illiquidity ratios, and measures reversal over a one-month horizon. It does not capture the co-movement of price and volume over a few days or deviations from moving averages at horizons of 3 to 24 trading days, which are the dimensions on which the robust Alpha191 signals load. This explanation is a conjecture consistent with the evidence, not a result of a formal test.

\subsection{Horizon Persistence}

A common concern with short-horizon signals is that their information decays within days, which would make monthly aggregation inappropriate. As a diagnostic, Online Appendix Table~A3 reports Newey-West $t$-statistics of the long-short returns over overlapping horizons of 1 to 21 trading days. None of the 32 signals that are significant at the one-day horizon loses significance at 21 days or in the scaled monthly series. The $t$-statistics generally increase with the horizon, but this should not be interpreted as a strengthening signal. With overlapping windows, Newey-West standard errors with $H$ lags understate the variance of the $H$-day mean, so the $t$-statistic rises by about 17\% at five days and 21\% at 21 days even when the daily premium is constant. The median observed increase is about 34\%, so most of it is mechanical. The evidence therefore supports the narrower conclusion that the signals' information does not decay within a month, which is what the monthly analysis requires. It does not show that the signals become stronger at longer horizons.

\newpage

\section{Discussion}\label{sec:discussion}

Our results show that part of the information in a trading-signal library built for Chinese retail-dominated markets is also relevant for pricing U.S.\ large caps. The results are statistical, so they cannot identify the economic mechanism, but two interpretations are consistent with the robust signals.

The first interpretation is liquidity demand combined with limits to arbitrage. Short-horizon price pressure from large or correlated trades can create temporary deviations from fundamental value, which arbitrageurs close only partly because of noise-trader risk and capital constraints \citep{delong1990noise, shleifer1997limits}. Mean-reversion signals such as 046 and 071 measure the distance of the price from its short-term moving averages. Signals such as 049 and 184 describe the direction and persistence of daily price pressure. Both types would naturally load on such dislocations. Behavioral explanations based on overreaction \citep{daniel1998investor, hirshleifer2001investor} lead to similar predictions. The short-term reversal literature \citep{jegadeesh1990evidence, lehmann1990fads} documents related effects at the monthly horizon. Our results suggest that the finer horizon structure captured by the Alpha191 formulas contains information beyond the monthly reversal factor included among our controls.

The second interpretation is information in order flow. In the model of \citet{kyle1985continuous}, informed traders split their orders over time, which leaves traces in the joint dynamics of price and volume. The volume-price correlation signals in our robust set (123, 001) and the On-Balance Volume and volume-MACD signals may proxy for this hidden demand. Under this interpretation, the signals reflect how information is incorporated into prices rather than irrational trading.

Our results do not establish that these mechanisms are universal. They show that a subset of signals designed for a retail-dominated market also contains information in the most institutionalized segment of the U.S.\ market, and that the transferable subset has a recognizable economic profile. Whether the common factor is investor behavior, market microstructure, or both remains an open question.

For practitioners, the scorecard approach is more useful than the list of baseline survivors. When hundreds of candidate signals are tested against hundreds of controls, the set of significant signals depends strongly on the test assets and on the method used to select controls. Requiring agreement across several reasonable specifications removes most candidates and leaves a small set that is more likely to reflect genuine structure.

\section{Limitations}\label{sec:limitations}

Several limitations qualify our conclusions and indicate directions for further work.

\textit{Implementation.} We do not evaluate turnover, transaction costs, or net-of-cost portfolio performance. The factors are monthly-rebalanced decile portfolios of S\&P 500 stocks, which limits trading costs compared with daily strategies. However, significant SDF loadings do not imply that a stand-alone strategy based on a signal is profitable after costs, and several robust signals have insignificant own returns.

\textit{Multiple testing and out-of-sample stability.} We test 168 signals, and only two meet a Bonferroni threshold in the baseline. The scorecard reduces the risk of false discoveries but is not a formal correction. We do not perform an out-of-sample, rolling-window, or subperiod analysis, so we cannot tell whether the loadings are stable across market regimes such as the 2008 financial crisis or the 2020 pandemic.

\textit{Scope and frequency.} The analysis covers S\&P 500 constituents only, a deliberate choice that provides a conservative test but leaves open whether the results extend to smaller stocks, where short-horizon effects are typically stronger. The monthly frequency is imposed by the control data. The horizon diagnostic indicates no decay within a month but does not measure the value of faster rebalancing.

\section{Conclusion}\label{sec:conclusion}

We test whether 168 trading signals developed for the Chinese A-share market add pricing information for S\&P 500 stocks beyond 153 established U.S.\ factors. A double-selection LASSO finds 17 signals with significant SDF loadings in the baseline design. A robustness scorecard reduces them to a core of three signals that survive all alternative specifications and six that survive most. The robust signals capture short-term mean reversion and volume-price dynamics, dimensions that the established U.S.\ factor zoo covers only partially, while volatility-based signals do not survive alternative specifications. The broader lesson for high-dimensional signal research is that significance in a single specification is a weak guide. A simple scorecard across test assets and control-selection methods separates stable signals from fragile ones at little additional cost.

\newpage
\bibliographystyle{plainnat}
\bibliography{references}

\clearpage
\setcounter{table}{0}
\renewcommand{\thetable}{A\arabic{table}}
\setcounter{page}{1}
\renewcommand{\thepage}{A-\arabic{page}}

\begin{center}
{\LARGE Online Appendix}\\[0.8em]
{\large to ``Cross-Market Alpha: Testing Short-Term Trading Factors in the U.S.\ Market via Double-Selection LASSO''}\\[0.5em]
{\normalsize Jin Du, Alexander Walter, Maxim Ulrich}
\end{center}
\vspace{1em}

\noindent This online appendix contains supplementary material to the main article:
\begin{itemize}
    \item Table~\ref{tab:A_5x5}: full DS-LASSO results for the $5\times5$ test-asset design.
    \item Table~\ref{tab:A_altmodels}: full Elastic Net and PCA benchmark results for the $3\times2$ design.
    \item Table~\ref{tab:A_horizon}: horizon diagnostic for all 168 signals.
    \item Table~\ref{tab:A_factors}: definitions and category assignments of the Alpha191 library.
\end{itemize}

\section*{A.1 \quad Full $5\times5$ DS-LASSO Results}

Table~\ref{tab:A_5x5} lists all Alpha191 signals with a significant SDF loading in the $5\times5$ (size $\times$ signal) design, next to the corresponding $3\times2$ estimates. $\lambda$ is the SDF loading on the covariance with the standardized factor, in basis points.

\setlength{\tabcolsep}{4pt}
\renewcommand{\arraystretch}{1.2}

\begin{longtable}{
    >{\raggedright\arraybackslash}p{0.43\textwidth}
    >{\centering\arraybackslash}p{0.12\textwidth}
    >{\centering\arraybackslash}p{0.12\textwidth}
    >{\centering\arraybackslash}p{0.12\textwidth}
    >{\centering\arraybackslash}p{0.12\textwidth}
}
\caption{DS-LASSO results, $5\times5$ vs.\ $3\times2$ test assets \label{tab:A_5x5}} \\
\toprule
~ & \multicolumn{2}{c}{Bivariate $5 \times 5$} & \multicolumn{2}{c}{Bivariate $3 \times 2$} \\
\cmidrule(lr){2-3} \cmidrule(lr){4-5}
Signal & $\lambda$ (bp) & $t$ & $\lambda$ (bp) & $t$ \\
\midrule
\endfirsthead
\caption*{Table \thetable{} -- continued} \\
\toprule
Signal & $\lambda$ (bp) & $t$ & $\lambda$ (bp) & $t$ \\
\midrule
\endhead
\midrule
\multicolumn{5}{r}{Continued on next page\ldots} \\
\endfoot
\bottomrule
\multicolumn{5}{l}{\footnotesize \textit{Note:} $^{*}$ and $^{**}$ indicate significance at the 5\% and 1\% levels, respectively.} \\
\endlastfoot
Rank of Delayed Price-Gap Correlation (184) & 40 & $6.78^{**}$ & 41 & $2.86^{**}$ \\
Inverse Rank of High-Price and Volume Correlation (141) & 25 & $5.54^{**}$ & 19 & $1.69$ \\
Price-Volume vs. Low-Volume Correlation Rank (123) & 26 & $5.43^{**}$ & 42 & $3.39^{**}$ \\
Overnight Gap Return (015) & 31 & $5.17^{**}$ & 41 & $2.46^{*}$ \\
Inverse Rank of Close-Volume Rank Covariance (099) & 22 & $5.04^{**}$ & 23 & $1.96$ \\
Multi-Period Mean Reversion Ratio (046) & 33 & $4.33^{**}$ & 79 & $3.68^{**}$ \\
Inverse Rank of Nested Decayed Price-Volume Correlations (073) & 21 & $4.29^{**}$ & 40 & $3.41^{**}$ \\
12-Day SMEA of Mean Deviation Momentum (022) & 34 & $4.24^{**}$ & -17 & $-0.86$ \\
Percentage Deviation from 12-Day Mean (031) & 20 & $4.22^{**}$ & 19 & $1.64$ \\
4-Day Rank of Weighted Price Change (006) & 27 & $4.03^{**}$ & 14 & $0.85$ \\
Volume-Weighted Momentum x Long Term Return Rank (025) & 24 & $3.94^{**}$ & 18 & $1.19$ \\
Volume MACD Histogram (155) & 20 & $3.67^{**}$ & 40 & $2.82^{**}$ \\
VWAP Momentum exponentiated by Volume Correlation (131) & 27 & $3.54^{**}$ & 25 & $1.26$ \\
6-Day Negative Correlation of Volume Growth and Return (001) & 20 & $3.41^{**}$ & 37 & $2.31^{*}$ \\
Downward Directional Pressure Ratio (049) & 11 & $3.41^{**}$ & 25 & $3.12^{**}$ \\
20-Day Cumulative Upward Gap Pressure (187) & 29 & $3.39^{**}$ & 25 & $1.11$ \\
Scaled 27/13-Day MACD Histogram (089) & 20 & $3.33^{**}$ & 22 & $1.41$ \\
Bullish-to-Bearish Typical Price Power (052) & 29 & $3.07^{**}$ & 35 & $1.36$ \\
1-Day Change in Intraday Price Position (002) & 16 & $2.93^{**}$ & 12 & $0.94$ \\
Combined VWAP Momentum and Low-Volume Corr (044) & 15 & $2.66^{**}$ & -2 & $-0.15$ \\
6-Day Cumulative Intraday Cash Flow Volume (011) & 22 & $2.58^{**}$ & -10 & $-0.48$ \\
Mean Reversion plus Long-Term VWAP-Price Corr (026) & 18 & $2.51^{*}$ & 1 & $0.07$ \\
Return Delta Rank x Open-Volume Correlation (136) & 13 & $2.50^{*}$ & 9 & $0.61$ \\
Complex Volume-Weighted Price Position Rank (170) & 19 & $2.50^{*}$ & 40 & $1.92$ \\
High-Price Exhaustion Delta (038) & 12 & $2.45^{*}$ & 6 & $0.51$ \\
VWAP Recovery vs. Volume Correlation Logic (154) & 12 & $2.36^{*}$ & 26 & $1.88$ \\
Volume-Low Corr Adjusted Mid-Price Spread (191) & 15 & $2.35^{*}$ & 34 & $1.83$ \\
Inverse Rank of Decayed Price-Volume Divergence (092) & 13 & $2.22^{*}$ & 26 & $1.75$ \\
20-Day Adjusted Price Movement Sum (059) & 16 & $2.13^{*}$ & 16 & $0.78$ \\
6-Day Average Directional Index (172) & 11 & $2.11^{*}$ & -10 & $-0.79$ \\
\end{longtable}
\endgroup
\endgroup

\bigskip
\section*{A.2 \quad Elastic Net and PCA Benchmarks}

Table~\ref{tab:A_altmodels} reports, for the 17 signals significant in the baseline $3\times2$ DS-LASSO, the corresponding estimates when the control set is chosen by cross-validated Elastic Net (ElasticNetCV over several $\ell_1$ ratios, followed by OLS on the selected controls plus all Alpha191 signals) or replaced by principal components of the control covariances that explain 90\% of their variance (followed by OLS on the components plus all Alpha191 signals). The $\lambda$ estimates are not directly comparable in magnitude to the DS estimates: the three procedures condition on different control sets, and the coefficients are SDF loadings conditional on those controls and on the remaining Alpha191 signals. The comparison of interest is the sign and significance pattern.

\begin{longtable}{
    >{\raggedright\arraybackslash}p{0.33\textwidth}
    *{6}{>{\centering\arraybackslash}p{0.09\textwidth}}
}
\caption{$3\times2$ results: DS-LASSO, Elastic Net, and PCA \label{tab:A_altmodels}} \\
\toprule
~ & \multicolumn{2}{c}{DS} & \multicolumn{2}{c}{ENet} & \multicolumn{2}{c}{PCA} \\
\cmidrule(lr){2-3} \cmidrule(lr){4-5} \cmidrule(lr){6-7}
Signal & $\lambda$ (bp) & $t$ & $\lambda$ (bp) & $t$ & $\lambda$ (bp) & $t$ \\
\midrule
\endfirsthead
\caption*{Table \thetable{} -- continued} \\
\toprule
Signal & $\lambda$ (bp) & $t$ & $\lambda$ (bp) & $t$ & $\lambda$ (bp) & $t$ \\
\midrule
\endhead
\midrule
\multicolumn{7}{r}{Continued on next page\ldots} \\
\endfoot
\bottomrule
\multicolumn{7}{l}{\footnotesize \textit{Note:} $^{*}$ and $^{**}$ indicate significance at the 5\% and 1\% levels, respectively.} \\
\endlastfoot
Multi-Period Mean Reversion Ratio (046) & 79 & $3.68^{**}$ & 678 & $4.34^{**}$ & 171 & $5.29^{**}$ \\
20-Day Cumulative On-Balance Volume (084) & 51 & $3.68^{**}$ & 449 & $4.43^{**}$ & 11 & $0.55$ \\
Inverse Rank of Nested Decayed Price-Volume Correlations (073) & 40 & $3.41^{**}$ & -266 & $-3.05^{**}$ & 28 & $2.10^{*}$ \\
Price-Volume vs. Low-Volume Correlation Rank (123) & 42 & $3.39^{**}$ & 57 & $1.15$ & 52 & $3.52^{**}$ \\
Downward Directional Pressure Ratio (049) & 25 & $3.12^{**}$ & 88 & $4.28^{**}$ & 12 & $2.10^{*}$ \\
24-Day Percentage Deviation from Mean (071) & 58 & $3.06^{**}$ & 685 & $4.00^{**}$ & 109 & $2.76^{**}$ \\
Rank of Delayed Price-Gap Correlation (184) & 41 & $2.86^{**}$ & 197 & $3.46^{**}$ & 72 & $5.06^{**}$ \\
Volume MACD Histogram (155) & 40 & $2.82^{**}$ & 116 & $2.02^{*}$ & -1 & $-0.05$ \\
Inverse Rank of Intraday Volatility and Corr (054) & 44 & $2.67^{**}$ & 471 & $3.52^{**}$ & 118 & $2.79^{**}$ \\
Benchmark-Relative Excess Return Skewness (181) & 38 & $2.57^{*}$ & -67 & $-1.38$ & 59 & $4.94^{**}$ \\
12-Day Average True Range (161) & 34 & $2.57^{*}$ & -1867 & $-3.94^{**}$ & 83 & $1.12$ \\
Log Gain-to-Loss Variability Ratio (190) & 36 & $2.49^{*}$ & -23 & $-0.38$ & 21 & $1.44$ \\
Rank of Decay-Adjusted Momentum-VWAP Divergence (039) & 39 & $2.49^{*}$ & 1 & $0.02$ & 20 & $1.29$ \\
Overnight Gap Return (015) & 41 & $2.46^{*}$ & 705 & $3.94^{**}$ & 39 & $1.02$ \\
6-Day Relative Strength Index (063) & 45 & $2.38^{*}$ & 326 & $2.53^{*}$ & -71 & $-2.94^{**}$ \\
6-Day Negative Correlation of Volume Growth and Return (001) & 37 & $2.31^{*}$ & 63 & $0.74$ & 41 & $2.33^{*}$ \\
10-Day Price Acceleration vs. Directional Change (086) & 32 & $2.30^{*}$ & 65 & $1.30$ & 7 & $0.47$ \\
\end{longtable}

\bigskip
\section*{A.3 \quad Horizon Diagnostic}

Table~\ref{tab:A_horizon} reports Newey-West adjusted $t$-statistics of the mean long-short return for overlapping $H$-day compounded returns, $H \in \{1,5,10,15,21\}$, computed with $H$ lags and a Bartlett kernel, together with the $t$-statistic of the scaled monthly return used in the main analysis (``Scaled Monthly''). The Spearman column is the rank correlation between $H$ and the absolute $t$-statistic; ``Trend'' labels series whose $t$-statistics rise monotonically in absolute value.

These statistics should be read as a persistence diagnostic, not as evidence that a signal strengthens with horizon. For overlapping sums of returns with a constant daily mean and no true autocorrelation, the Bartlett-kernel estimator with $H$ lags understates the variance of the $H$-day mean return; the implied $t$-statistic is mechanically larger than the one-day value by a factor of roughly 1.17 at $H=5$, 1.20 at $H=10$ and $H=15$, and 1.21 at $H=21$. Rising $t$-statistics are therefore expected even in the absence of any horizon structure. Across the 75 signals with $|t|>1$ at $H=1$, the median ratio of the 21-day to the one-day $t$-statistic is 1.34; most of this increase is accounted for by the mechanical effect, and the remainder is consistent with mild positive autocorrelation in the daily long-short returns. The informative feature of the table is the absence of decay: signals that are significant at the one-day horizon remain significant at the monthly horizon. Signals 018/019 and 050/051 produce identical rows because their definitions yield the same cross-sectional ranking.

\begingroup\small
\setlength{\tabcolsep}{4pt}
\renewcommand{\arraystretch}{1.05}
\begin{longtable}{lcccccccl}
\caption{Newey-West $t$-statistics by return horizon \label{tab:A_horizon}}\\
\toprule
\multirow{2}{*}{Signal} & \multicolumn{6}{c}{Horizon $H$ (days)} & \multirow{2}{*}{Spearman} & \multirow{2}{*}{Trend} \\
\cmidrule(lr){2-7}
 & 1 & 5 & 10 & 15 & 21 & Scaled Monthly & & \\
\midrule
\endfirsthead
\caption*{Table \thetable{} -- continued} \\
\toprule
\multirow{2}{*}{Signal} & \multicolumn{6}{c}{Horizon $H$ (days)} & \multirow{2}{*}{Spearman} & \multirow{2}{*}{Trend} \\
\cmidrule(lr){2-7}
 & 1 & 5 & 10 & 15 & 21 & Scaled Monthly & & \\
\midrule
\endhead
\endfoot
\bottomrule
\endlastfoot
Alpha\_001 & 0.29 & 0.59 & 0.58 & 0.59 & 0.56 & 0.37 & 0.10 & Mixed \\
Alpha\_002 & 2.33 & 2.77 & 2.89 & 2.96 & 3.06 & 2.98 & 1.00 & Monotonic Increase \\
Alpha\_003 & -2.49 & -2.99 & -3.21 & -3.33 & -3.45 & -2.85 & 1.00 & Monotonic Increase \\
Alpha\_004 & 0.31 & 0.38 & 0.38 & 0.39 & 0.39 & 0.32 & 0.90 & Mixed \\
Alpha\_006 & 0.92 & 1.13 & 1.18 & 1.24 & 1.26 & 1.05 & 1.00 & Monotonic Increase \\
Alpha\_007 & -0.62 & -0.76 & -0.99 & -1.11 & -1.16 & -0.81 & 1.00 & Monotonic Increase \\
Alpha\_008 & -0.52 & -0.61 & -0.70 & -0.72 & -0.70 & -0.63 & 0.70 & Mixed \\
Alpha\_009 & -2.14 & -2.83 & -3.06 & -3.09 & -3.04 & -2.31 & 0.70 & Mixed \\
Alpha\_010 & -1.97 & -2.44 & -2.59 & -2.66 & -2.72 & -2.31 & 1.00 & Monotonic Increase \\
Alpha\_011 & -2.19 & -2.65 & -2.82 & -2.86 & -2.90 & -2.41 & 1.00 & Monotonic Increase \\
Alpha\_012 & 2.14 & 2.57 & 2.63 & 2.68 & 2.70 & 2.30 & 1.00 & Monotonic Increase \\
Alpha\_013 & 2.52 & 3.06 & 3.15 & 3.21 & 3.25 & 2.82 & 1.00 & Monotonic Increase \\
Alpha\_014 & -0.04 & -0.06 & -0.10 & -0.14 & -0.16 & -0.07 & 1.00 & Monotonic Increase \\
Alpha\_015 & 2.49 & 3.02 & 3.14 & 3.20 & 3.24 & 2.71 & 1.00 & Monotonic Increase \\
Alpha\_017 & -0.32 & -0.39 & -0.47 & -0.50 & -0.55 & -0.35 & 1.00 & Monotonic Increase \\
Alpha\_018 & -1.03 & -1.33 & -1.53 & -1.66 & -1.88 & -1.45 & 1.00 & Monotonic Increase \\
Alpha\_019 & -1.03 & -1.33 & -1.53 & -1.66 & -1.88 & -1.45 & 1.00 & Monotonic Increase \\
Alpha\_020 & -0.43 & -0.56 & -0.72 & -0.79 & -0.85 & -0.61 & 1.00 & Monotonic Increase \\
Alpha\_021 & -0.48 & -0.53 & -0.67 & -0.73 & -0.77 & -0.59 & 1.00 & Monotonic Increase \\
Alpha\_022 & -0.74 & -0.95 & -1.05 & -1.15 & -1.37 & -1.14 & 1.00 & Monotonic Increase \\
Alpha\_023 & 1.14 & 1.37 & 1.41 & 1.43 & 1.43 & 1.38 & 0.90 & Mixed \\
Alpha\_024 & -0.69 & -0.83 & -0.98 & -1.03 & -1.05 & -0.79 & 1.00 & Monotonic Increase \\
Alpha\_025 & -0.25 & -0.26 & -0.32 & -0.34 & -0.37 & -0.32 & 1.00 & Monotonic Increase \\
Alpha\_026 & 0.05 & 0.11 & 0.04 & 0.03 & -0.01 & 0.08 & -0.90 & Mixed \\
Alpha\_027 & -0.37 & -0.48 & -0.54 & -0.60 & -0.68 & -0.56 & 1.00 & Monotonic Increase \\
Alpha\_028 & -2.12 & -2.56 & -2.78 & -2.85 & -2.87 & -2.31 & 1.00 & Monotonic Increase \\
Alpha\_029 & -0.57 & -0.69 & -0.83 & -0.93 & -1.04 & -0.75 & 1.00 & Monotonic Increase \\
Alpha\_031 & -0.40 & -0.50 & -0.64 & -0.68 & -0.71 & -0.52 & 1.00 & Monotonic Increase \\
Alpha\_032 & 0.65 & 0.75 & 0.74 & 0.74 & 0.72 & 0.74 & 0.10 & Mixed \\
Alpha\_033 & 0.81 & 1.04 & 1.11 & 1.12 & 1.14 & 1.14 & 1.00 & Monotonic Increase \\
Alpha\_034 & 0.40 & 0.54 & 0.47 & 0.48 & 0.50 & 0.52 & 0.40 & Mixed \\
Alpha\_035 & -1.73 & -2.30 & -2.61 & -2.65 & -2.58 & -1.84 & 0.70 & Mixed \\
Alpha\_037 & 0.05 & 0.04 & 0.00 & -0.00 & -0.02 & 0.09 & -0.60 & Mixed \\
Alpha\_038 & 0.54 & 0.64 & 0.64 & 0.62 & 0.60 & 0.53 & 0.00 & Mixed \\
Alpha\_039 & -1.03 & -1.32 & -1.39 & -1.42 & -1.42 & -1.08 & 1.00 & Monotonic Increase \\
Alpha\_040 & -0.47 & -0.59 & -0.68 & -0.73 & -0.81 & -0.51 & 1.00 & Monotonic Increase \\
Alpha\_041 & -0.64 & -0.77 & -0.89 & -0.89 & -0.89 & -0.76 & 1.00 & Monotonic Increase \\
Alpha\_042 & -0.67 & -0.94 & -1.07 & -1.11 & -1.12 & -0.87 & 1.00 & Monotonic Increase \\
Alpha\_043 & -0.04 & -0.07 & -0.19 & -0.24 & -0.34 & -0.13 & 1.00 & Monotonic Increase \\
Alpha\_044 & 0.36 & 0.44 & 0.43 & 0.43 & 0.42 & 0.45 & 0.00 & Mixed \\
Alpha\_045 & 0.77 & 0.87 & 0.86 & 0.84 & 0.82 & 0.87 & 0.00 & Mixed \\
Alpha\_046 & 0.18 & 0.28 & 0.17 & 0.16 & 0.16 & 0.21 & -0.90 & Mixed \\
Alpha\_047 & 1.69 & 2.08 & 2.17 & 2.21 & 2.23 & 1.96 & 1.00 & Monotonic Increase \\
Alpha\_048 & -0.53 & -0.58 & -0.64 & -0.68 & -0.72 & -0.60 & 1.00 & Monotonic Increase \\
Alpha\_049 & 0.14 & 0.21 & 0.07 & 0.05 & 0.03 & 0.22 & -0.90 & Mixed \\
Alpha\_050 & -0.14 & -0.20 & -0.38 & -0.45 & -0.51 & -0.22 & 1.00 & Monotonic Increase \\
Alpha\_051 & -0.14 & -0.20 & -0.38 & -0.45 & -0.51 & -0.22 & 1.00 & Monotonic Increase \\
Alpha\_052 & -2.49 & -3.08 & -3.21 & -3.25 & -3.31 & -2.81 & 1.00 & Monotonic Increase \\
Alpha\_053 & -0.63 & -0.80 & -0.91 & -1.00 & -1.07 & -0.71 & 1.00 & Monotonic Increase \\
Alpha\_054 & -2.51 & -3.06 & -3.25 & -3.32 & -3.37 & -2.78 & 1.00 & Monotonic Increase \\
Alpha\_055 & -2.12 & -2.52 & -2.66 & -2.70 & -2.72 & -2.24 & 1.00 & Monotonic Increase \\
Alpha\_056 & 0.70 & 0.89 & 0.90 & 0.90 & 0.91 & 0.69 & 0.90 & Mixed \\
Alpha\_057 & -1.70 & -2.09 & -2.28 & -2.35 & -2.40 & -1.95 & 1.00 & Monotonic Increase \\
Alpha\_058 & 0.52 & 0.61 & 0.60 & 0.60 & 0.59 & 0.76 & 0.10 & Mixed \\
Alpha\_059 & -2.53 & -3.03 & -3.25 & -3.34 & -3.42 & -2.80 & 1.00 & Monotonic Increase \\
Alpha\_060 & -2.05 & -2.45 & -2.59 & -2.61 & -2.63 & -2.21 & 1.00 & Monotonic Increase \\
Alpha\_061 & -0.03 & 0.04 & -0.01 & -0.02 & -0.01 & -0.03 & -0.60 & Mixed \\
Alpha\_062 & 0.09 & 0.20 & 0.16 & 0.14 & 0.16 & 0.10 & 0.30 & Mixed \\
Alpha\_063 & -0.56 & -0.69 & -0.92 & -1.00 & -1.07 & -0.71 & 1.00 & Monotonic Increase \\
Alpha\_065 & 0.99 & 1.28 & 1.32 & 1.33 & 1.33 & 1.33 & 1.00 & Monotonic Increase \\
Alpha\_066 & -0.99 & -1.30 & -1.57 & -1.69 & -1.77 & -1.33 & 1.00 & Monotonic Increase \\
Alpha\_067 & 0.50 & 0.66 & 0.57 & 0.56 & 0.49 & 0.68 & -0.40 & Mixed \\
Alpha\_068 & -2.44 & -3.44 & -4.01 & -4.14 & -4.21 & -3.20 & 1.00 & Monotonic Increase \\
Alpha\_069 & 0.44 & 0.60 & 0.55 & 0.54 & 0.52 & 0.56 & 0.00 & Mixed \\
Alpha\_071 & 0.21 & 0.37 & 0.25 & 0.21 & 0.11 & 0.30 & -0.70 & Mixed \\
Alpha\_072 & 1.57 & 1.93 & 2.00 & 2.04 & 2.05 & 1.77 & 1.00 & Monotonic Increase \\
Alpha\_073 & 0.55 & 0.67 & 0.64 & 0.66 & 0.64 & 0.76 & 0.10 & Mixed \\
Alpha\_076 & 1.26 & 1.54 & 1.56 & 1.58 & 1.59 & 1.59 & 1.00 & Monotonic Increase \\
Alpha\_077 & 1.85 & 2.27 & 2.35 & 2.39 & 2.40 & 2.00 & 1.00 & Monotonic Increase \\
Alpha\_078 & -1.18 & -1.65 & -1.93 & -2.01 & -2.08 & -1.56 & 1.00 & Monotonic Increase \\
Alpha\_079 & -0.67 & -0.76 & -0.91 & -0.97 & -1.00 & -0.75 & 1.00 & Monotonic Increase \\
Alpha\_080 & 0.30 & 0.46 & 0.38 & 0.34 & 0.28 & 0.41 & -0.40 & Mixed \\
Alpha\_081 & -2.03 & -2.55 & -2.78 & -2.91 & -2.97 & -2.53 & 1.00 & Monotonic Increase \\
Alpha\_082 & 1.60 & 1.96 & 2.03 & 2.06 & 2.07 & 1.76 & 1.00 & Monotonic Increase \\
Alpha\_084 & 0.46 & 0.52 & 0.50 & 0.48 & 0.46 & 0.61 & 0.00 & Mixed \\
Alpha\_085 & 0.33 & 0.50 & 0.46 & 0.43 & 0.39 & 0.45 & 0.00 & Mixed \\
Alpha\_086 & 2.46 & 2.91 & 3.07 & 3.08 & 3.06 & 2.51 & 0.70 & Mixed \\
Alpha\_087 & -1.06 & -1.30 & -1.36 & -1.36 & -1.35 & -1.05 & 0.60 & Mixed \\
Alpha\_088 & 1.32 & 1.67 & 1.70 & 1.75 & 1.75 & 1.71 & 0.90 & Mixed \\
Alpha\_089 & -1.16 & -1.42 & -1.66 & -1.85 & -1.98 & -1.52 & 1.00 & Monotonic Increase \\
Alpha\_091 & 1.71 & 2.09 & 2.15 & 2.19 & 2.24 & 1.95 & 1.00 & Monotonic Increase \\
Alpha\_092 & 0.63 & 0.82 & 0.84 & 0.87 & 0.87 & 0.69 & 1.00 & Monotonic Increase \\
Alpha\_093 & 0.73 & 0.85 & 0.84 & 0.82 & 0.81 & 0.73 & 0.00 & Mixed \\
Alpha\_094 & 0.96 & 1.17 & 1.21 & 1.23 & 1.21 & 1.13 & 0.70 & Mixed \\
Alpha\_096 & -1.75 & -2.15 & -2.35 & -2.44 & -2.51 & -2.05 & 1.00 & Monotonic Increase \\
Alpha\_097 & -1.43 & -1.85 & -2.05 & -2.15 & -2.20 & -1.80 & 1.00 & Monotonic Increase \\
Alpha\_098 & 1.10 & 1.36 & 1.42 & 1.46 & 1.46 & 1.34 & 1.00 & Monotonic Increase \\
Alpha\_099 & 1.31 & 1.61 & 1.74 & 1.80 & 1.82 & 1.56 & 1.00 & Monotonic Increase \\
Alpha\_100 & -1.79 & -2.30 & -2.54 & -2.65 & -2.71 & -2.25 & 1.00 & Monotonic Increase \\
Alpha\_102 & 0.65 & 0.86 & 0.90 & 0.87 & 0.86 & 0.78 & 0.60 & Mixed \\
Alpha\_103 & -0.40 & -0.48 & -0.55 & -0.62 & -0.68 & -0.53 & 1.00 & Monotonic Increase \\
Alpha\_104 & -1.09 & -1.36 & -1.43 & -1.45 & -1.45 & -1.24 & 1.00 & Monotonic Increase \\
Alpha\_105 & 0.11 & 0.19 & 0.04 & -0.04 & -0.13 & 0.11 & -0.20 & Mixed \\
Alpha\_106 & -0.17 & -0.25 & -0.33 & -0.35 & -0.39 & -0.12 & 1.00 & Monotonic Increase \\
Alpha\_107 & -1.11 & -1.34 & -1.49 & -1.53 & -1.56 & -1.32 & 1.00 & Monotonic Increase \\
Alpha\_108 & 0.95 & 1.18 & 1.26 & 1.28 & 1.28 & 1.06 & 0.90 & Mixed \\
Alpha\_109 & -0.29 & -0.31 & -0.33 & -0.32 & -0.33 & -0.31 & 0.90 & Mixed \\
Alpha\_111 & -0.68 & -0.76 & -0.79 & -0.80 & -0.81 & -0.68 & 1.00 & Monotonic Increase \\
Alpha\_112 & -0.40 & -0.49 & -0.67 & -0.84 & -0.97 & -0.56 & 1.00 & Monotonic Increase \\
Alpha\_113 & -0.84 & -0.98 & -1.10 & -1.19 & -1.33 & -1.08 & 1.00 & Monotonic Increase \\
Alpha\_114 & -1.06 & -1.30 & -1.49 & -1.53 & -1.55 & -1.28 & 1.00 & Monotonic Increase \\
Alpha\_116 & -1.39 & -1.75 & -2.04 & -2.16 & -2.23 & -1.60 & 1.00 & Monotonic Increase \\
Alpha\_118 & 0.73 & 0.91 & 0.94 & 0.94 & 0.92 & 0.88 & 0.60 & Mixed \\
Alpha\_119 & -1.29 & -1.65 & -1.77 & -1.82 & -1.91 & -1.58 & 1.00 & Monotonic Increase \\
Alpha\_120 & 1.70 & 2.06 & 2.09 & 2.09 & 2.08 & 1.69 & 0.60 & Mixed \\
Alpha\_122 & 0.50 & 0.73 & 0.65 & 0.64 & 0.57 & 0.77 & 0.00 & Mixed \\
Alpha\_123 & -0.18 & -0.23 & -0.26 & -0.28 & -0.28 & -0.18 & 1.00 & Monotonic Increase \\
Alpha\_124 & 0.85 & 1.10 & 1.05 & 1.01 & 1.01 & 1.03 & 0.10 & Mixed \\
Alpha\_125 & 2.51 & 2.95 & 3.29 & 3.42 & 3.39 & 2.83 & 0.90 & Mixed \\
Alpha\_126 & -2.29 & -2.64 & -2.71 & -2.70 & -2.77 & -2.43 & 0.90 & Mixed \\
Alpha\_128 & -0.85 & -1.07 & -1.18 & -1.28 & -1.35 & -1.00 & 1.00 & Monotonic Increase \\
Alpha\_129 & -1.87 & -2.24 & -2.40 & -2.51 & -2.61 & -2.20 & 1.00 & Monotonic Increase \\
Alpha\_131 & 0.77 & 0.95 & 0.93 & 0.88 & 0.88 & 0.90 & 0.00 & Mixed \\
Alpha\_133 & 0.26 & 0.37 & 0.31 & 0.25 & 0.23 & 0.31 & -0.70 & Mixed \\
Alpha\_134 & -0.55 & -0.66 & -0.78 & -0.84 & -0.91 & -0.66 & 1.00 & Monotonic Increase \\
Alpha\_135 & 0.95 & 1.18 & 1.14 & 1.16 & 1.16 & 1.36 & 0.30 & Mixed \\
Alpha\_136 & -0.95 & -1.20 & -1.30 & -1.35 & -1.41 & -1.06 & 1.00 & Monotonic Increase \\
Alpha\_137 & -2.47 & -2.98 & -3.16 & -3.24 & -3.30 & -2.73 & 1.00 & Monotonic Increase \\
Alpha\_138 & 0.31 & 0.45 & 0.38 & 0.38 & 0.39 & 0.37 & 0.30 & Mixed \\
Alpha\_139 & -0.98 & -1.26 & -1.35 & -1.36 & -1.38 & -1.00 & 1.00 & Monotonic Increase \\
Alpha\_141 & 1.52 & 1.86 & 2.03 & 2.17 & 2.23 & 2.11 & 1.00 & Monotonic Increase \\
Alpha\_142 & -0.25 & -0.28 & -0.39 & -0.47 & -0.57 & -0.27 & 1.00 & Monotonic Increase \\
Alpha\_143 & -0.53 & -0.58 & -0.65 & -0.70 & -0.76 & -0.60 & 1.00 & Monotonic Increase \\
Alpha\_145 & 0.39 & 0.47 & 0.46 & 0.46 & 0.45 & 0.41 & 0.00 & Mixed \\
Alpha\_146 & 1.23 & 1.45 & 1.48 & 1.50 & 1.51 & 1.46 & 1.00 & Monotonic Increase \\
Alpha\_147 & -0.78 & -0.96 & -1.12 & -1.21 & -1.28 & -0.86 & 1.00 & Monotonic Increase \\
Alpha\_148 & 1.29 & 1.55 & 1.65 & 1.71 & 1.71 & 1.49 & 1.00 & Monotonic Increase \\
Alpha\_149 & 1.12 & 1.35 & 1.36 & 1.38 & 1.39 & 1.37 & 1.00 & Monotonic Increase \\
Alpha\_150 & -2.11 & -2.66 & -2.80 & -2.87 & -2.93 & -2.55 & 1.00 & Monotonic Increase \\
Alpha\_151 & 0.72 & 0.87 & 0.85 & 0.85 & 0.83 & 0.86 & 0.00 & Mixed \\
Alpha\_152 & 0.71 & 0.87 & 0.87 & 0.86 & 0.86 & 0.97 & 0.00 & Mixed \\
Alpha\_153 & -3.03 & -3.80 & -3.93 & -3.97 & -4.05 & -3.41 & 1.00 & Monotonic Increase \\
Alpha\_154 & 0.78 & 0.97 & 1.01 & 1.02 & 1.03 & 0.82 & 1.00 & Monotonic Increase \\
Alpha\_155 & 0.58 & 0.72 & 0.74 & 0.73 & 0.74 & 0.64 & 0.90 & Mixed \\
Alpha\_156 & -0.04 & 0.01 & -0.05 & -0.03 & -0.02 & -0.06 & -0.20 & Mixed \\
Alpha\_157 & 0.81 & 1.00 & 1.03 & 1.06 & 1.08 & 0.89 & 1.00 & Monotonic Increase \\
Alpha\_158 & 3.83 & 4.70 & 4.78 & 4.76 & 4.80 & 3.91 & 0.90 & Mixed \\
Alpha\_159 & -2.11 & -2.55 & -2.68 & -2.71 & -2.73 & -2.33 & 1.00 & Monotonic Increase \\
Alpha\_160 & -2.16 & -2.67 & -2.89 & -3.00 & -3.10 & -2.55 & 1.00 & Monotonic Increase \\
Alpha\_161 & -2.66 & -3.22 & -3.34 & -3.35 & -3.39 & -2.91 & 1.00 & Monotonic Increase \\
Alpha\_162 & -1.19 & -1.60 & -1.88 & -1.95 & -1.98 & -1.49 & 1.00 & Monotonic Increase \\
Alpha\_163 & 0.19 & 0.27 & 0.26 & 0.23 & 0.17 & 0.23 & -0.40 & Mixed \\
Alpha\_164 & 1.56 & 1.88 & 1.95 & 1.96 & 1.97 & 1.74 & 1.00 & Monotonic Increase \\
Alpha\_166 & 0.14 & 0.19 & 0.06 & 0.02 & -0.03 & 0.21 & -0.80 & Mixed \\
Alpha\_167 & -2.64 & -3.22 & -3.40 & -3.51 & -3.63 & -2.97 & 1.00 & Monotonic Increase \\
Alpha\_168 & -1.01 & -1.20 & -1.49 & -1.67 & -1.75 & -1.36 & 1.00 & Monotonic Increase \\
Alpha\_169 & 0.00 & 0.05 & -0.05 & -0.11 & -0.13 & 0.10 & 1.00 & Monotonic Increase \\
Alpha\_170 & 1.47 & 1.83 & 1.89 & 1.93 & 1.98 & 1.70 & 1.00 & Monotonic Increase \\
Alpha\_171 & 1.30 & 1.56 & 1.61 & 1.63 & 1.64 & 1.37 & 1.00 & Monotonic Increase \\
Alpha\_172 & -0.96 & -1.22 & -1.34 & -1.38 & -1.48 & -1.25 & 1.00 & Monotonic Increase \\
Alpha\_173 & -2.77 & -3.41 & -3.54 & -3.61 & -3.71 & -3.20 & 1.00 & Monotonic Increase \\
Alpha\_174 & -1.99 & -2.47 & -2.63 & -2.70 & -2.78 & -2.27 & 1.00 & Monotonic Increase \\
Alpha\_175 & -2.74 & -3.33 & -3.47 & -3.49 & -3.52 & -3.03 & 1.00 & Monotonic Increase \\
Alpha\_176 & -1.53 & -2.05 & -2.37 & -2.53 & -2.65 & -2.12 & 1.00 & Monotonic Increase \\
Alpha\_177 & 0.11 & 0.20 & 0.13 & 0.09 & 0.08 & 0.10 & -0.70 & Mixed \\
Alpha\_178 & -0.76 & -0.87 & -1.06 & -1.21 & -1.31 & -1.05 & 1.00 & Monotonic Increase \\
Alpha\_179 & -0.42 & -0.43 & -0.55 & -0.60 & -0.63 & -0.37 & 1.00 & Monotonic Increase \\
Alpha\_180 & 1.29 & 1.67 & 1.75 & 1.75 & 1.74 & 1.49 & 0.60 & Mixed \\
Alpha\_181 & 1.26 & 1.55 & 1.65 & 1.69 & 1.71 & 1.82 & 1.00 & Monotonic Increase \\
Alpha\_182 & 0.23 & 0.31 & 0.32 & 0.33 & 0.34 & 0.22 & 1.00 & Monotonic Increase \\
Alpha\_184 & 1.42 & 1.66 & 1.73 & 1.75 & 1.77 & 1.55 & 1.00 & Monotonic Increase \\
Alpha\_185 & 1.28 & 1.46 & 1.44 & 1.41 & 1.40 & 1.33 & 0.00 & Mixed \\
Alpha\_186 & -0.37 & -0.39 & -0.49 & -0.53 & -0.59 & -0.42 & 1.00 & Monotonic Increase \\
Alpha\_187 & 0.97 & 1.11 & 1.12 & 1.12 & 1.11 & 0.95 & 0.30 & Mixed \\
Alpha\_188 & -1.32 & -1.57 & -1.75 & -1.81 & -1.84 & -1.61 & 1.00 & Monotonic Increase \\
Alpha\_189 & -2.34 & -2.87 & -3.08 & -3.19 & -3.27 & -2.67 & 1.00 & Monotonic Increase \\
Alpha\_190 & 0.14 & 0.23 & 0.16 & 0.18 & 0.18 & 0.14 & 0.40 & Mixed \\
Alpha\_191 & 2.18 & 2.63 & 2.73 & 2.79 & 2.83 & 2.35 & 1.00 & Monotonic Increase \\
\end{longtable}
\endgroup

\newpage
\begin{landscape}
\section*{A.4 \quad The Alpha191 Library}

Definitions follow the original Guotai Junan Securities specification. The category column is our heuristic assignment to six thematic groups. Signals marked in the Ex.\ column (23 in total) were excluded before estimation because their time series had excessive missing values or were numerically unstable in the rolling-window computation; the remaining 168 signals are used in the analysis.

\setlength{\tabcolsep}{4pt}
\renewcommand{\arraystretch}{1.1}
\newcolumntype{L}[1]{>{\raggedright\arraybackslash}p{#1}}
\newcolumntype{C}[1]{>{\centering\arraybackslash}p{#1}}
\begin{longtable}[l]{L{1.8cm} L{2.7cm} L{5.0cm} C{0.8cm} L{13.9cm}}
\caption{Alpha191 signal definitions and categories \label{tab:A_factors}}\\
\toprule
\textbf{Signal} & \textbf{Category} & \textbf{Long Name} & \textbf{Ex.} & \textbf{Formula} \\
\midrule
\endfirsthead
\caption*{Table \thetable{} -- continued}\\
\toprule
\textbf{Signal} & \textbf{Category} & \textbf{Long Name} & \textbf{Ex.} & \textbf{Formula} \\
\midrule
\endhead
\midrule
\multicolumn{5}{r}{Continued on next page\ldots} \\
\endfoot
\bottomrule
\endlastfoot
\small 001 & \small Volume \& Flow & \small 6-Day Negative Correlation of Volume Growth and Return &  & \fml{(-1*CORR(RANK(DELTA(LOG(VOLUME), 1)), RANK(((CLOSE - OPEN) / OPEN)), 6))} \\
\small 002 & \small Mean Reversion & \small 1-Day Change in Intraday Price Position &  & \fml{(-1 * DELTA((((CLOSE - LOW) - (HIGH - CLOSE)) / (HIGH - LOW)), 1))} \\
\small 003 & \small Trend \& Momentum & \small 6-Day Cumulative Price Movement vs. Range Extremes &  & \fml{SUM((CLOSE=DELAY(CLOSE,1)?0:CLOSE-(CLOSE>DELAY(CLOSE,1)?MIN(LOW,DELAY(CLOSE,1)):MAX(HIGH,DELAY(CLOSE,1)))),6)} \\
\small 004 & \small Volatility \& Risk & \small Bollinger Band Breakout with Volume Confirmation &  & \fml{((((SUM(CLOSE, 8) / 8) + STD(CLOSE, 8)) < (SUM(CLOSE, 2) / 2)) ? (-1 * 1) : (((SUM(CLOSE, 2) / 2) < ((SUM(CLOSE, 8) / 8) - STD(CLOSE, 8))) ? 1 : (((1 < (VOLUME / MEAN(VOLUME,20))) || ((VOLUME / MEAN(VOLUME,20)) == 1)) ? 1 : (-1 * 1))))} \\
\small 005 & \small Volume \& Flow & \small 3-Day Max Negative Correlation of Volume and High Ranks & x & \fml{(-1 * TSMAX(CORR(TSRANK(VOLUME, 5), TSRANK(HIGH, 5), 5), 3))} \\
\small 006 & \small Trend \& Momentum & \small 4-Day Rank of Weighted Price Change &  & \fml{(RANK(SIGN(DELTA((((OPEN * 0.85) + (HIGH * 0.15))), 4)))* -1)} \\
\small 007 & \small Liquidity \& VWAP & \small VWAP Distance Range weighted by Volume Change &  & \fml{((RANK(MAX((VWAP - CLOSE), 3)) + RANK(MIN((VWAP - CLOSE), 3))) * RANK(DELTA(VOLUME, 3)))} \\
\small 008 & \small Trend \& Momentum & \small 4-Day Rank of Negative Mid-Price Momentum &  & \fml{RANK(DELTA(((((HIGH + LOW) / 2) * 0.2) + (VWAP * 0.8)), 4) * -1)} \\
\small 009 & \small Liquidity \& VWAP & \small 7-Day SMA of Price Impact per Volume Unit &  & \fml{SMA(((HIGH+LOW)/2-(DELAY(HIGH,1)+DELAY(LOW,1))/2)*(HIGH-LOW)/VOLUME,7,2)} \\
\small 010 & \small Volatility \& Risk & \small 5-Day Rank of Max Volatility-Adjusted Returns &  & \fml{(RANK(MAX(((RET < 0) ? STD(RET, 20) : CLOSE)^2),5))} \\
\small 011 & \small Volume \& Flow & \small 6-Day Cumulative Intraday Cash Flow Volume &  & \fml{SUM(((CLOSE-LOW)-(HIGH-CLOSE))./(HIGH-LOW).*VOLUME,6)} \\
\small 012 & \small Liquidity \& VWAP & \small Inverse Rank of Open-VWAP spread vs. VWAP Deviation &  & \fml{(RANK((OPEN - (SUM(VWAP, 10) / 10)))) * (-1 * (RANK(ABS((CLOSE - VWAP)))))} \\
\small 013 & \small Mean Reversion & \small Geometric Mean Mid-Price to VWAP Spread &  & \fml{(((HIGH * LOW)^0.5) - VWAP)} \\
\small 014 & \small Trend \& Momentum & \small 5-Day Absolute Price Change &  & \fml{CLOSE-DELAY(CLOSE,5)} \\
\small 015 & \small Trend \& Momentum & \small Overnight Gap Return &  & \fml{OPEN/DELAY(CLOSE,1)-1} \\
\small 016 & \small Volume \& Flow & \small 5-Day Max Inverse Rank of Volume-VWAP Correlation & x & \fml{(-1 * TSMAX(RANK(CORR(RANK(VOLUME), RANK(VWAP), 5)), 5))} \\
\small 017 & \small Liquidity \& VWAP & \small VWAP Distance to High exponentiated by 5-Day Delta &  & \fml{RANK((VWAP - MAX(VWAP, 15)))^DELTA(CLOSE, 5)} \\
\small 018 & \small Trend \& Momentum & \small 5-Day Price Ratio &  & \fml{CLOSE/DELAY(CLOSE,5)} \\
\small 019 & \small Trend \& Momentum & \small Conditional 5-Day Percentage Change &  & \fml{(CLOSE<DELAY(CLOSE,5)?(CLOSE-DELAY(CLOSE,5))/DELAY(CLOSE,5):(CLOSE=DELAY(CLOSE,5)?0:(CLOSE-DELAY(CLOSE,5))/CLOSE))} \\
\small 020 & \small Trend \& Momentum & \small 6-Day Percentage Return &  & \fml{(CLOSE-DELAY(CLOSE,6))/DELAY(CLOSE,6)*100} \\
\small 021 & \small Trend \& Momentum & \small 6-Day Linear Regression Trend Slope &  & \fml{REGBETA(MEAN(CLOSE,6),SEQUENCE(6))} \\
\small 022 & \small Mean Reversion & \small 12-Day SMEA of Mean Deviation Momentum &  & \fml{SMEAN(((CLOSE-MEAN(CLOSE,6))/MEAN(CLOSE,6)-DELAY((CLOSE-MEAN(CLOSE,6))/MEAN(CLOSE,6),3)),12,1)} \\
\small 023 & \small Volatility \& Risk & \small Ratio of Upside Volatility to Total Volatility &  & \fml{SMA((CLOSE>DELAY(CLOSE,1)?STD(CLOSE:20),0),20,1)/(SMA((CLOSE>DELAY(CLOSE,1)?STD(CLOSE,20):0),20,1)+SMA((CLOSE<=DELAY(CLOSE,1)?STD(CLOSE,20):0),20,1))*100} \\
\small 024 & \small Trend \& Momentum & \small 5-Day SMA of Price Momentum &  & \fml{SMA(CLOSE-DELAY(CLOSE,5),5,1)} \\
\small 025 & \small Trend \& Momentum & \small Volume-Weighted Momentum x Long Term Return Rank &  & \fml{((-1 * RANK((DELTA(CLOSE, 7) * (1 - RANK(DECAYLINEAR((VOLUME / MEAN(VOLUME,20)), 9)))))) * (1 + RANK(SUM(RET, 250))))} \\
\small 026 & \small Mean Reversion & \small Mean Reversion plus Long-Term VWAP-Price Corr &  & \fml{((((SUM(CLOSE, 7) / 7) - CLOSE)) + ((CORR(VWAP, DELAY(CLOSE, 5), 230))))} \\
\small 027 & \small Trend \& Momentum & \small 12-Day WMA of Combined Price Changes &  & \fml{WMA((CLOSE-DELAY(CLOSE,3))/DELAY(CLOSE,3)*100+(CLOSE-DELAY(CLOSE,6))/DELAY(CLOSE,6)*100,12)} \\
\small 028 & \small Mean Reversion & \small Triple-Smoothed Stochastic Oscillator Logic &  & \fml{3*SMA((CLOSE-TSMIN(LOW,9))/(TSMAX(HIGH,9)-TSMIN(LOW,9))*100,3,1)-2*SMA(SMA((CLOSE-TSMIN(LOW,9))/(MAX(HIGH,9)-TSMAX(LOW,9))*100,3,1),3,1)} \\
\small 029 & \small Volume \& Flow & \small Volume-Weighted 6-Day Momentum &  & \fml{(CLOSE-DELAY(CLOSE,6))/DELAY(CLOSE,6)*VOLUME} \\
\small 030 & \small Volatility \& Risk & \small 20-Day WMA of Fama-French Residual Variance & x & \fml{WMA((REGRESI(CLOSE/DELAY(CLOSE)-1,MKT,SMB,HML,60))^2,20)} \\
\small 031 & \small Mean Reversion & \small Percentage Deviation from 12-Day Mean &  & \fml{(CLOSE-MEAN(CLOSE,12))/MEAN(CLOSE,12)*100} \\
\small 032 & \small Volume \& Flow & \small 3-Day Sum of Rank Volume-High Correlation &  & \fml{(-1 * SUM(RANK(CORR(RANK(HIGH), RANK(VOLUME), 3)), 3))} \\
\small 033 & \small Trend \& Momentum & \small 5-Day Low Change x Long Term Relative Momentum &  & \fml{((((-1 * TSMIN(LOW, 5)) + DELAY(TSMIN(LOW, 5), 5)) * RANK(((SUM(RET, 240) - SUM(RET, 20)) / 220))) * TSRANK(VOLUME, 5))} \\
\small 034 & \small Mean Reversion & \small 12-Day Price-to-Mean Ratio &  & \fml{MEAN(CLOSE,12)/CLOSE} \\
\small 035 & \small Volume \& Flow & \small Inverse Rank of Decay-Adjusted Open-Volume Flow &  & \fml{(MIN(RANK(DECAYLINEAR(DELTA(OPEN, 1), 15)), RANK(DECAYLINEAR(CORR((VOLUME), ((OPEN * 0.65) + (OPEN *0.35)), 17),7))) * -1)} \\
\small 036 & \small Volume \& Flow & \small 2-Day Rank of 6-Day Volume-VWAP Correlation & x & \fml{RANK(SUM(CORR(RANK(VOLUME), RANK(VWAP)), 6), 2)} \\
\small 037 & \small Trend \& Momentum & \small Rank of Change in 5-Day Open-Return Product &  & \fml{(-1 * RANK(((SUM(OPEN, 5) * SUM(RET, 5)) - DELAY((SUM(OPEN, 5) * SUM(RET, 5)), 10))))} \\
\small 038 & \small Mean Reversion & \small High-Price Exhaustion Delta &  & \fml{(((SUM(HIGH, 20) / 20) < HIGH) ? (-1 * DELTA(HIGH, 2)) : 0)} \\
\small 039 & \small Liquidity \& VWAP & \small Rank of Decay-Adjusted Momentum-VWAP Divergence &  & \fml{((RANK(DECAYLINEAR(DELTA((CLOSE), 2),8)) - RANK(DECAYLINEAR(CORR(((VWAP * 0.3) + (OPEN * 0.7)), SUM(MEAN(VOLUME,180), 37), 14), 12))) * -1)} \\
\small 040 & \small Volume \& Flow & \small 26-Day Up-Volume to Down-Volume Ratio &  & \fml{SUM((CLOSE>DELAY(CLOSE,1)?VOLUME:0),26)/SUM((CLOSE<=DELAY(CLOSE,1)?VOLUME:0),26)*100} \\
\small 041 & \small Liquidity \& VWAP & \small Inverse Rank of 5-Day Max VWAP Momentum &  & \fml{(RANK(MAX(DELTA((VWAP), 3), 5))* -1)} \\
\small 042 & \small Price Action & \small High-Price Volatility x Volume Correlation &  & \fml{((-1 * RANK(STD(HIGH, 10))) * CORR(HIGH, VOLUME, 10))} \\
\small 043 & \small Volume \& Flow & \small 6-Day Chaikin-style Money Flow &  & \fml{SUM((CLOSE>DELAY(CLOSE,1)?VOLUME:(CLOSE<DELAY(CLOSE,1)?-VOLUME:0)),6)} \\
\small 044 & \small Liquidity \& VWAP & \small Combined VWAP Momentum and Low-Volume Corr &  & \fml{(TSRANK(DECAYLINEAR(CORR(((LOW )), MEAN(VOLUME,10), 7), 6),4) + TSRANK(DECAYLINEAR(DELTA((VWAP), 3), 10), 15))} \\
\small 045 & \small Volume \& Flow & \small Price Change x VWAP-Volume Correlation Rank &  & \fml{(RANK(DELTA((((CLOSE * 0.6) + (OPEN *0.4))), 1)) * RANK(CORR(VWAP, MEAN(VOLUME,150), 15)))} \\
\small 046 & \small Mean Reversion & \small Multi-Period Mean Reversion Ratio &  & \fml{(MEAN(CLOSE,3)+MEAN(CLOSE,6)+MEAN(CLOSE,12)+MEAN(CLOSE,24))/(4*CLOSE)} \\
\small 047 & \small Mean Reversion & \small 9-Day SMA of Williams \% Reversed &  & \fml{SMA((TSMAX(HIGH,6)-CLOSE)/(TSMAX(HIGH,6)-TSMIN(LOW,6))*100,9,1)} \\
\small 048 & \small Trend \& Momentum & \small Volume-Weighted 3-Day Price Direction Rank &  & \fml{(-1*((RANK(((SIGN((CLOSE - DELAY(CLOSE, 1))) + SIGN((DELAY(CLOSE, 1) - DELAY(CLOSE, 2)))) + SIGN((DELAY(CLOSE, 2) - DELAY(CLOSE, 3)))))) * SUM(VOLUME, 5)) / SUM(VOLUME, 20))} \\
\small 049 & \small Trend \& Momentum & \small Downward Directional Pressure Ratio &  & \fml{SUM(((HIGH+LOW)>=(DELAY(HIGH,1)+DELAY(LOW,1))?0:MAX(ABS(HIGH-DELAY(HIGH,1)),ABS(LOW-DELAY(LOW,1)))),12)/(SUM(((HIGH+LOW)>=(DELAY(HIGH,1)+DELAY(LOW,1))?0:MAX(ABS(HIGH-DELAY(HIGH,1)),ABS(LOW-DELAY(LOW,1)))),12)+SUM(((HIGH+LOW)<=(DELAY(HIGH,1)+DELAY(LOW,1))?0:MAX(ABS(HIGH-DELAY(HIGH,1)),ABS(LOW-DELAY(LOW,1)))),12))} \\
\small 050 & \small Trend \& Momentum & \small Net Directional Movement Index &  & \fml{SUM(((HIGH+LOW)<=(DELAY(HIGH,1)+DELAY(LOW,1))?0:MAX(ABS(HIGH-DELAY(HIGH,1)),ABS(LOW-DELAY(LOW,1)))),12)/(SUM(((HIGH+LOW)<=(DELAY(HIGH,1)+DELAY(LOW,1))?0:MAX(ABS(HIGH-DELAY(HIGH,1)),ABS(LOW-DELAY(LOW,1)))),12)+SUM(((HIGH+LOW)>=(DELAY(HIGH,1)+DELAY(LOW,1))?0:MAX(ABS(HIGH-DELAY(HIGH,1)),ABS(LOW-DELAY(LOW,1)))),12))-SUM(((HIGH+LOW)>=(DELAY(HIGH,1)+DELAY(LOW,1))?0:MAX(ABS(HIGH-DELAY(HIGH,1)),ABS(LOW-DELAY(LOW,1)))),12)/(SUM(((HIGH+LOW)>=(DELAY(HIGH,1)+DELAY(LOW,1))?0:MAX(ABS(HIGH-DELAY(HIGH,1)),ABS(LOW-DELAY(LOW,1)))),12)+SUM(((HIGH+LOW)<=(DELAY(HIGH,1)+DELAY(LOW,1))?0:MAX(ABS(HIGH-DELAY(HIGH,1)),ABS(LOW-DELAY(LOW,1)))),12))} \\
\small 051 & \small Trend \& Momentum & \small Upward Directional Pressure Ratio &  & \fml{SUM(((HIGH+LOW)<=(DELAY(HIGH,1)+DELAY(LOW,1))?0:MAX(ABS(HIGH-DELAY(HIGH,1)),ABS(LOW-DELAY(LOW,1)))),12)/(SUM(((HIGH+LOW)<=(DELAY(HIGH,1)+DELAY(LOW,1))?0:MAX(ABS(HIGH-DELAY(HIGH,1)),ABS(LOW-DELAY(LOW,1)))),12)+SUM(((HIGH+LOW)>=(DELAY(HIGH,1)+DELAY(LOW,1))?0:MAX(ABS(HIGH-DELAY(HIGH,1)),ABS(LOW-DELAY(LOW,1)))),12))} \\
\small 052 & \small Trend \& Momentum & \small Bullish-to-Bearish Typical Price Power &  & \fml{SUM(MAX(0,HIGH-DELAY((HIGH+LOW+CLOSE)/3,1)),26)/SUM(MAX(0,DELAY((HIGH+LOW+CLOSE)/3,1)-L),26)*100} \\
\small 053 & \small Trend \& Momentum & \small 12-Day Up-Day Percentage Frequency &  & \fml{COUNT(CLOSE>DELAY(CLOSE,1),12)/12*100} \\
\small 054 & \small Volatility \& Risk & \small Inverse Rank of Intraday Volatility and Corr &  & \fml{(-1 * RANK((STD(ABS(CLOSE - OPEN)) + (CLOSE - OPEN)) + CORR(CLOSE, OPEN,10)))} \\
\small 055 & \small Trend \& Momentum & \small 20-Day Cumulative Accumulation Swing Index &  & \fml{SUM(16*(CLOSE-DELAY(CLOSE,1)+(CLOSE-OPEN)/2+DELAY(CLOSE,1)-DELAY(OPEN,1))/((ABS(HIGH-DELAY(CLOSE,1))>ABS(LOW-DELAY(CLOSE,1)) \& ABS(HIGH-DELAY(CLOSE,1))>ABS(HIGH-DELAY(LOW,1))?ABS(HIGH-DELAY(CLOSE,1))+ABS(LOW-DELAY(CLOSE,1))/2+ABS(DELAY(CLOSE,1)-DELAY(OPEN,1))/4:(ABS(LOW-DELAY(CLOSE,1))>ABS(HIGH-DELAY(LOW,1)) \& ABS(LOW-DELAY(CLOSE,1))>ABS(HIGH-DELAY(CLOSE,1))?ABS(LOW-DELAY(CLOSE,1))+ABS(HIGH-DELAY(CLOSE,1))/2+ABS(DELAY(CLOSE,1)-DELAY(OPEN,1))/4:ABS(HIGH-DELAY(LOW,1))+ABS(DELAY(CLOSE,1)-DELAY(OPEN,1))/4)))*MAX(ABS(HIGH-DELAY(CLOSE,1)),ABS(LOW-DELAY(CLOSE,1))),20)} \\
\small 056 & \small Mean Reversion & \small Relative Rank of Open-Price Recovery &  & \fml{(RANK((OPEN - TSMIN(OPEN, 12))) < RANK((RANK(CORR(SUM(((HIGH + LOW) / 2), 19), SUM(MEAN(VOLUME,40), 19), 13))^5)))} \\
\small 057 & \small Mean Reversion & \small 3-Day SMA of K-Stochastic Logic &  & \fml{SMA((CLOSE-TSMIN(LOW,9))/(TSMAX(HIGH,9)-TSMIN(LOW,9))*100,3,1)} \\
\small 058 & \small Trend \& Momentum & \small 20-Day Up-Day Percentage Frequency &  & \fml{COUNT(CLOSE>DELAY(CLOSE,1),20)/20*100} \\
\small 059 & \small Trend \& Momentum & \small 20-Day Adjusted Price Movement Sum &  & \fml{SUM((CLOSE=DELAY(CLOSE,1)?0:CLOSE-(CLOSE>DELAY(CLOSE,1)?MIN(LOW,DELAY(CLOSE,1)):MAX(HIGH,DELAY(CLOSE,1)))),20)} \\
\small 060 & \small Volume \& Flow & \small 20-Day Intraday Accumulation/Distribution &  & \fml{SUM(((CLOSE-LOW)-(HIGH-CLOSE))./(HIGH-LOW).*VOLUME,20)} \\
\small 061 & \small Volume \& Flow & \small Inverse Rank of Decayed VWAP Momentum and Volume Correlation &  & \fml{(MAX(RANK(DECAYLINEAR(DELTA(VWAP, 1), 12)), RANK(DECAYLINEAR(RANK(CORR((LOW),MEAN(VOLUME,80), 8)), 17))) * -1)} \\
\small 062 & \small Volume \& Flow & \small 5-Day Negative Correlation of High Price and Volume Rank &  & \fml{(-1 * CORR(HIGH, RANK(VOLUME), 5))} \\
\small 063 & \small Trend \& Momentum & \small 6-Day Relative Strength Index &  & \fml{SMA(MAX(CLOSE-DELAY(CLOSE,1),0),6,1)/SMA(ABS(CLOSE-DELAY(CLOSE,1)),6,1)*100} \\
\small 064 & \small Volume \& Flow & \small Inverse Rank of Decayed VWAP-Volume and Close-Volume Corrs & x & \fml{(MAX(RANK(DECAYLINEAR(CORR(RANK(VWAP), RANK(VOLUME), 4), 4)), RANK(DECAYLINEAR(MAX(CORR(RANK(CLOSE), RANK(MEAN(VOLUME,60)), 4), 13), 14))) * -1)} \\
\small 065 & \small Mean Reversion & \small 6-Day Price-to-Mean Ratio &  & \fml{MEAN(CLOSE,6)/CLOSE} \\
\small 066 & \small Mean Reversion & \small 6-Day Percentage Deviation from Mean &  & \fml{(CLOSE-MEAN(CLOSE,6))/MEAN(CLOSE,6)*100} \\
\small 067 & \small Trend \& Momentum & \small 24-Day Relative Strength Index &  & \fml{SMA(MAX(CLOSE-DELAY(CLOSE,1),0),24,1)/SMA(ABS(CLOSE-DELAY(CLOSE,1)),24,1)*100} \\
\small 068 & \small Liquidity \& VWAP & \small 15-Day SMA of Price Impact per Volume Unit &  & \fml{SMA(((HIGH+LOW)/2-(DELAY(HIGH,1)+DELAY(LOW,1))/2)*(HIGH-LOW)/VOLUME,15,2)} \\
\small 069 & \small Trend \& Momentum & \small 20-Day Directional Typical Price Movement Ratio &  & \fml{(SUM(DTM,20)>SUM(DBM,20)?(SUM(DTM,20)-SUM(DBM,20))/SUM(DTM,20):(SUM(DTM,20)=SUM(DBM,20)?0:(SUM(DTM,20)-SUM(DBM,20))/SUM(DBM,20)))} \\
\small 070 & \small Volatility \& Risk & \small 6-Day Standard Deviation of Transaction Amount & x & \fml{STD(AMOUNT,6)} \\
\small 071 & \small Mean Reversion & \small 24-Day Percentage Deviation from Mean &  & \fml{(CLOSE-MEAN(CLOSE,24))/MEAN(CLOSE,24)*100} \\
\small 072 & \small Mean Reversion & \small 15-Day SMA of Williams \% Reversed &  & \fml{SMA((TSMAX(HIGH,6)-CLOSE)/(TSMAX(HIGH,6)-TSMIN(LOW,6))*100,15,1)} \\
\small 073 & \small Volume \& Flow & \small Inverse Rank of Nested Decayed Price-Volume Correlations &  & \fml{((TSRANK(DECAYLINEAR(DECAYLINEAR(CORR((CLOSE), VOLUME, 10), 16), 4), 5) -RANK(DECAYLINEAR(CORR(VWAP, MEAN(VOLUME,30), 4),3))) * -1)} \\
\small 074 & \small Volume \& Flow & \small Combined Volume-Price Rank Correlations & x & \fml{(RANK(CORR(SUM(((LOW * 0.35) + (VWAP * 0.65)), 20), SUM(MEAN(VOLUME,40), 20), 7)) + RANK(CORR(RANK(VWAP), RANK(VOLUME), 6)))} \\
\small 075 & \small Trend \& Momentum & \small Benchmark-Relative Outperformance Frequency & x & \fml{COUNT(CLOSE>OPEN \& BANCHMARKINDEXCLOSE<BANCHMARKINDEXOPEN,50)/COUNT(BANCHMARKINDEXCLOSE<BANCHMARKINDEXOPEN,50)} \\
\small 076 & \small Liquidity \& VWAP & \small 20-Day Volatility-to-Volume Efficiency Ratio &  & \fml{STD(ABS((CLOSE/DELAY(CLOSE,1)-1))/VOLUME,20)/MEAN(ABS((CLOSE/DELAY(CLOSE,1)-1))/VOLUME,20)} \\
\small 077 & \small Volume \& Flow & \small Min Rank of Decayed Price Spread and Vol Correlation &  & \fml{MIN(RANK(DECAYLINEAR(((((HIGH + LOW) / 2) + HIGH) - (VWAP + HIGH)), 20)), RANK(DECAYLINEAR(CORR(((HIGH + LOW) / 2), MEAN(VOLUME,40), 3), 6)))} \\
\small 078 & \small Mean Reversion & \small Commodity Channel Index Logic &  & \fml{((HIGH+LOW+CLOSE)/3-MA((HIGH+LOW+CLOSE)/3,12))/(0.015*MEAN(ABS(CLOSE-MEAN((HIGH+LOW+CLOSE)/3,12)),12))} \\
\small 079 & \small Trend \& Momentum & \small 12-Day Relative Strength Index &  & \fml{SMA(MAX(CLOSE-DELAY(CLOSE,1),0),12,1)/SMA(ABS(CLOSE-DELAY(CLOSE,1)),12,1)*100} \\
\small 080 & \small Volume \& Flow & \small 5-Day Volume Percentage Change &  & \fml{(VOLUME-DELAY(VOLUME,5))/DELAY(VOLUME,5)*100} \\
\small 081 & \small Volume \& Flow & \small 21-Day Simple Moving Average of Volume &  & \fml{SMA(VOLUME,21,2)} \\
\small 082 & \small Mean Reversion & \small 20-Day SMA of Williams \% Reversed &  & \fml{SMA((TSMAX(HIGH,6)-CLOSE)/(TSMAX(HIGH,6)-TSMIN(LOW,6))*100,20,1)} \\
\small 083 & \small Volume \& Flow & \small Inverse Rank of High-Volume Rank Covariance & x & \fml{(-1 * RANK(COVIANCE(RANK(HIGH), RANK(VOLUME), 5)))} \\
\small 084 & \small Volume \& Flow & \small 20-Day Cumulative On-Balance Volume &  & \fml{SUM((CLOSE>DELAY(CLOSE,1)?VOLUME:(CLOSE<DELAY(CLOSE,1)?-VOLUME:0)),20)} \\
\small 085 & \small Volume \& Flow & \small Volume Rank x 7-Day Negative Return Rank &  & \fml{(TSRANK((VOLUME / MEAN(VOLUME,20)), 20) * TSRANK((-1 * DELTA(CLOSE, 7)), 8))} \\
\small 086 & \small Trend \& Momentum & \small 10-Day Price Acceleration vs. Directional Change &  & \fml{((0.25 < (((DELAY(CLOSE, 20) - DELAY(CLOSE, 10)) / 10) - ((DELAY(CLOSE, 10) - CLOSE) / 10))) ? (-1 * 1) : (((((DELAY(CLOSE, 20) - DELAY(CLOSE, 10)) / 10) - ((DELAY(CLOSE, 10) - CLOSE) / 10)) < 0) ? 1 : ((-1 * 1) * (CLOSE - DELAY(CLOSE, 1)))))} \\
\small 087 & \small Liquidity \& VWAP & \small Combined Decayed VWAP Momentum and Spread Rank &  & \fml{((RANK(DECAYLINEAR(DELTA(VWAP, 4), 7)) + TSRANK(DECAYLINEAR(((((LOW * 0.9) + (LOW * 0.1)) - VWAP) / (OPEN - ((HIGH + LOW) / 2))), 11), 7)) * -1)} \\
\small 088 & \small Trend \& Momentum & \small 20-Day Percentage Return &  & \fml{(CLOSE-DELAY(CLOSE,20))/DELAY(CLOSE,20)*100} \\
\small 089 & \small Trend \& Momentum & \small Scaled 27/13-Day MACD Histogram &  & \fml{2*(SMA(CLOSE,13,2)-SMA(CLOSE,27,2)-SMA(SMA(CLOSE,13,2)-SMA(CLOSE,27,2),10,2))} \\
\small 090 & \small Volume \& Flow & \small Negative Rank of VWAP-Volume Correlation & x & \fml{( RANK(CORR(RANK(VWAP), RANK(VOLUME), 5)) * -1)} \\
\small 091 & \small Volume \& Flow & \small Price Deviation x Volume-Low Correlation Rank &  & \fml{((RANK((CLOSE - MAX(CLOSE, 5)))*RANK(CORR((MEAN(VOLUME,40)), LOW, 5))) * -1)} \\
\small 092 & \small Liquidity \& VWAP & \small Inverse Rank of Decayed Price-Volume Divergence &  & \fml{(MAX(RANK(DECAYLINEAR(DELTA(((CLOSE * 0.35) + (VWAP *0.65)), 2), 3)), TSRANK(DECAYLINEAR(ABS(CORR((MEAN(VOLUME,180)), CLOSE, 13)), 5), 15)) * -1)} \\
\small 093 & \small Volatility \& Risk & \small 20-Day Cumulative Downward Gap Pressure &  & \fml{SUM((OPEN>=DELAY(OPEN,1)?0:MAX((OPEN-LOW),(OPEN-DELAY(OPEN,1)))),20)} \\
\small 094 & \small Volume \& Flow & \small 30-Day Cumulative On-Balance Volume &  & \fml{SUM((CLOSE>DELAY(CLOSE,1)?VOLUME:(CLOSE<DELAY(CLOSE,1)?-VOLUME:0)),30)} \\
\small 095 & \small Volatility \& Risk & \small 20-Day Standard Deviation of Transaction Amount & x & \fml{STD(AMOUNT,20)} \\
\small 096 & \small Mean Reversion & \small 9-Day Double-Smoothed Stochastic &  & \fml{SMA(SMA((CLOSE-TSMIN(LOW,9))/(TSMAX(HIGH,9)-TSMIN(LOW,9))*100,3,1),3,1)} \\
\small 097 & \small Volatility \& Risk & \small 10-Day Standard Deviation of Volume &  & \fml{STD(VOLUME,10)} \\
\small 098 & \small Trend \& Momentum & \small 100-Day Trend Break vs. 3-Day Delta &  & \fml{((((DELTA((SUM(CLOSE, 100) / 100), 100) / DELAY(CLOSE, 100)) < 0.05) || ((DELTA((SUM(CLOSE, 100) / 100), 100) / DELAY(CLOSE, 100)) == 0.05)) ? (-1 * (CLOSE - TSMIN(CLOSE, 100))) : (-1 * DELTA(CLOSE, 3)))} \\
\small 099 & \small Volume \& Flow & \small Inverse Rank of Close-Volume Rank Covariance &  & \fml{(-1 * RANK(COVIANCE(RANK(CLOSE), RANK(VOLUME), 5)))} \\
\small 100 & \small Volatility \& Risk & \small 20-Day Standard Deviation of Volume &  & \fml{STD(VOLUME,20)} \\
\small 101 & \small Volume \& Flow & \small Relative Rank of VWAP-Volume vs. Price-Volume Correlation & x & \fml{((RANK(CORR(CLOSE, SUM(MEAN(VOLUME,30), 37), 15)) < RANK(CORR(RANK(((HIGH * 0.1) + (VWAP * 0.9))), RANK(VOLUME), 11))) * -1)} \\
\small 102 & \small Volume \& Flow & \small 6-Day Volume RSI &  & \fml{SMA(MAX(VOLUME-DELAY(VOLUME,1),0),6,1)/SMA(ABS(VOLUME-DELAY(VOLUME,1)),6,1)*100} \\
\small 103 & \small Trend \& Momentum & \small 20-Day Low-Day Recency &  & \fml{((20-LOWDAY(LOW,20))/20)*100} \\
\small 104 & \small Price Action & \small Volume Correlation Change x Volatility Rank &  & \fml{(-1 * (DELTA(CORR(HIGH, VOLUME, 5), 5) * RANK(STD(CLOSE, 20))))} \\
\small 105 & \small Volume \& Flow & \small Inverse Rank of Open-Volume Correlation &  & \fml{(-1 * CORR(RANK(OPEN), RANK(VOLUME), 10))} \\
\small 106 & \small Trend \& Momentum & \small 20-Day Absolute Price Change &  & \fml{CLOSE-DELAY(CLOSE,20)} \\
\small 107 & \small Trend \& Momentum & \small Combined Rank of Opening Gap Positions &  & \fml{(((-1 * RANK((OPEN - DELAY(HIGH, 1)))) * RANK((OPEN - DELAY(CLOSE, 1)))) * RANK((OPEN - DELAY(LOW, 1))))} \\
\small 108 & \small Liquidity \& VWAP & \small High-Price Breakthrough x VWAP-Volume Rank &  & \fml{((RANK((HIGH - MIN(HIGH, 2)))^RANK(CORR((VWAP), (MEAN(VOLUME,120)), 6))) * -1)} \\
\small 109 & \small Volatility \& Risk & \small 10-Day SMA Ratio of Price Range &  & \fml{SMA(HIGH-LOW,10,2)/SMA(SMA(HIGH-LOW,10,2),10,2)} \\
\small 110 & \small Trend \& Momentum & \small Bull/Bear Range Power Ratio & x & \fml{SUM(MAX(0,HIGH-DELAY(CLOSE,1)),20)/SUM(MAX(0,DELAY(CLOSE,1)-LOW),20)*100} \\
\small 111 & \small Volume \& Flow & \small Intraday Money Flow Oscillator &  & \fml{SMA(VOL*((CLOSE-LOW)-(HIGH-CLOSE))/(HIGH-LOW),11,2)-SMA(VOL*((CLOSE-LOW)-(HIGH-CLOSE))/(HIGH-LOW),4,2)} \\
\small 112 & \small Trend \& Momentum & \small 12-Day Chande Momentum Oscillator &  & \fml{(SUM((CLOSE-DELAY(CLOSE,1)>0?CLOSE-DELAY(CLOSE,1):0),12)-SUM((CLOSE-DELAY(CLOSE,1)<0?ABS(CLOSE-DELAY(CLOSE,1)):0),12))/(SUM((CLOSE-DELAY(CLOSE,1)>0?CLOSE-DELAY(CLOSE,1):0),12)+SUM((CLOSE-DELAY(CLOSE,1)<0?ABS(CLOSE-DELAY(CLOSE,1)):0),12))*100} \\
\small 113 & \small Volume \& Flow & \small Price-Volume Rank Correlation x Mean Return &  & \fml{(-1 * ((RANK((SUM(DELAY(CLOSE, 5), 20) / 20)) * CORR(CLOSE, VOLUME, 2)) * RANK(CORR(SUM(CLOSE, 5), SUM(CLOSE, 20), 2))))} \\
\small 114 & \small Liquidity \& VWAP & \small Volume Rank x Volatility Efficiency Ratio &  & \fml{((RANK(DELAY(((HIGH - LOW) / (SUM(CLOSE, 5) / 5)), 2)) * RANK(RANK(VOLUME))) / (((HIGH - LOW) / (SUM(CLOSE, 5) / 5)) / (VWAP - CLOSE)))} \\
\small 115 & \small Volume \& Flow & \small Nested Rank of Price-Volume and Range-Volume Corrs & x & \fml{(RANK(CORR(((HIGH * 0.9) + (CLOSE * 0.1)), MEAN(VOLUME,30), 10))^RANK(CORR(TSRANK(((HIGH + LOW) / 2), 4), TSRANK(VOLUME, 10), 7)))} \\
\small 116 & \small Trend \& Momentum & \small 20-Day Linear Regression Beta &  & \fml{REGBETA(CLOSE,SEQUENCE,20)} \\
\small 117 & \small Trend \& Momentum & \small Triple Rank: Volume, Price Position, and Return & x & \fml{((TSRANK(VOLUME, 32) * (1 - TSRANK(((CLOSE + HIGH) - LOW), 16))) * (1 - TSRANK(RET, 32)))} \\
\small 118 & \small Price Action & \small Upside vs. Downside Shadow Ratio &  & \fml{SUM(HIGH-OPEN,20)/SUM(OPEN-LOW,20)*100} \\
\small 119 & \small Volume \& Flow & \small Decayed VWAP-Volume vs. Open-Volume Rank Divergence &  & \fml{(RANK(DECAYLINEAR(CORR(VWAP, SUM(MEAN(VOLUME,5), 26), 5), 7)) -RANK(DECAYLINEAR(TSRANK(MIN(CORR(RANK(OPEN), RANK(MEAN(VOLUME,15)), 21), 9), 7), 8)))} \\
\small 120 & \small Mean Reversion & \small VWAP-to-Close Relative Position Rank &  & \fml{(RANK((VWAP - CLOSE)) / RANK((VWAP + CLOSE)))} \\
\small 121 & \small Liquidity \& VWAP & \small Inverse Rank of VWAP Recovery x Vol-VWAP Corr & x & \fml{((RANK((VWAP - MIN(VWAP, 12)))^TSRANK(CORR(TSRANK(VWAP, 20), TSRANK(MEAN(VOLUME,60), 2), 18), 3)) * -1)} \\
\small 122 & \small Trend \& Momentum & \small Triple Exponential Average Growth Rate &  & \fml{(SMA(SMA(SMA(LOG(CLOSE),13,2),13,2),13,2)-DELAY(SMA(SMA(SMA(LOG(CLOSE),13,2),13,2),13,2),1))/DELAY(SMA(SMA(SMA(LOG(CLOSE),13,2),13,2),13,2),1)} \\
\small 123 & \small Volume \& Flow & \small Price-Volume vs. Low-Volume Correlation Rank &  & \fml{((RANK(CORR(SUM(((HIGH + LOW) / 2), 20), SUM(MEAN(VOLUME,60), 20), 9)) < RANK(CORR(LOW, VOLUME, 6))) * -1)} \\
\small 124 & \small Mean Reversion & \small VWAP Distance Adjusted by Max Price Rank &  & \fml{(CLOSE - VWAP) / DECAYLINEAR(RANK(TSMAX(CLOSE, 30)),2)} \\
\small 125 & \small Liquidity \& VWAP & \small Decayed VWAP-Volume Corr vs. Momentum Rank &  & \fml{(RANK(DECAYLINEAR(CORR((VWAP), MEAN(VOLUME,80),17), 20)) / RANK(DECAYLINEAR(DELTA(((CLOSE * 0.5) + (VWAP * 0.5)), 3), 16)))} \\
\small 126 & \small Mean Reversion & \small Typical Price &  & \fml{(CLOSE+HIGH+LOW)/3} \\
\small 127 & \small Volatility \& Risk & \small RMS of 12-Day Price Drawdown & x & \fml{(MEAN((100*(CLOSE-MAX(CLOSE,12))/(MAX(CLOSE,12)))^2))^(1/2)} \\
\small 128 & \small Volume \& Flow & \small 14-Day Money Flow Index &  & \fml{100-(100/(1+SUM(((HIGH+LOW+CLOSE)/3>DELAY((HIGH+LOW+CLOSE)/3,1)?(HIGH+LOW+CLOSE)/3*VOLUME:0),14)/SUM(((HIGH+LOW+CLOSE)/3<DELAY((HIGH+LOW+CLOSE)/3,1)?(HIGH+LOW+CLOSE)/3*VOLUME:0),14)))} \\
\small 129 & \small Trend \& Momentum & \small 12-Day Cumulative Downside Absolute Change &  & \fml{SUM((CLOSE-DELAY(CLOSE,1)<0?ABS(CLOSE-DELAY(CLOSE,1)):0),12)} \\
\small 130 & \small Volume \& Flow & \small Decayed Mid-Price Vol Corr vs. VWAP-Vol Rank & x & \fml{(RANK(DECAYLINEAR(CORR(((HIGH + LOW) / 2), MEAN(VOLUME,40), 9), 10)) / RANK(DECAYLINEAR(CORR(RANK(VWAP), RANK(VOLUME), 7),3)))} \\
\small 131 & \small Liquidity \& VWAP & \small VWAP Momentum exponentiated by Volume Correlation &  & \fml{(RANK(DELAT(VWAP, 1))^TSRANK(CORR(CLOSE,MEAN(VOLUME,50), 18), 18))} \\
\small 132 & \small Liquidity \& VWAP & \small 20-Day Average Transaction Amount & x & \fml{MEAN(AMOUNT,20)} \\
\small 133 & \small Trend \& Momentum & \small 20-Day Aroon Oscillator &  & \fml{((20-HIGHDAY(HIGH,20))/20)*100-((20-LOWDAY(LOW,20))/20)*100} \\
\small 134 & \small Volume \& Flow & \small 12-Day Volume-Weighted Momentum &  & \fml{(CLOSE-DELAY(CLOSE,12))/DELAY(CLOSE,12)*VOLUME} \\
\small 135 & \small Trend \& Momentum & \small 20-Day SMA of 20-Day Price Ratios &  & \fml{SMA(DELAY(CLOSE/DELAY(CLOSE,20),1),20,1)} \\
\small 136 & \small Volume \& Flow & \small Return Delta Rank x Open-Volume Correlation &  & \fml{((-1 * RANK(DELTA(RET, 3))) * CORR(OPEN, VOLUME, 10))} \\
\small 137 & \small Price Action & \small Single-Day Accumulation Swing Index &  & \fml{16*(CLOSE-DELAY(CLOSE,1)+(CLOSE-OPEN)/2+DELAY(CLOSE,1)-DELAY(OPEN,1))/((ABS(HIGH-DELAY(CLOSE,1))>ABS(LOW-DELAY(CLOSE,1)) \& ABS(HIGH-DELAY(CLOSE,1))>ABS(HIGH-DELAY(LOW,1))?ABS(HIGH-DELAY(CLOSE,1))+ABS(LOW-DELAY(CLOSE,1))/2+ABS(DELAY(CLOSE,1)-DELAY(OPEN,1))/4:(ABS(LOW-DELAY(CLOSE,1))>ABS(HIGH-DELAY(LOW,1)) \& ABS(LOW-DELAY(CLOSE,1))>ABS(HIGH-DELAY(CLOSE,1))?ABS(LOW-DELAY(CLOSE,1))+ABS(HIGH-DELAY(CLOSE,1))/2+ABS(DELAY(CLOSE,1)-DELAY(OPEN,1))/4:ABS(HIGH-DELAY(LOW,1))+ABS(DELAY(CLOSE,1)-DELAY(OPEN,1))/4)))*MAX(ABS(HIGH-DELAY(CLOSE,1)),ABS(LOW-DELAY(CLOSE,1)))} \\
\small 138 & \small Trend \& Momentum & \small Inverse Rank of Decayed Low Momentum vs. Vol Corr &  & \fml{((RANK(DECAYLINEAR(DELTA((((LOW * 0.7) + (VWAP *0.3))), 3), 20)) -TSRANK(DECAYLINEAR(TSRANK(CORR(TSRANK(LOW, 8), TSRANK(MEAN(VOLUME,60), 17), 5), 19), 16), 7)) * -1)} \\
\small 139 & \small Volume \& Flow & \small Negative Open-Volume Correlation &  & \fml{(-1 * CORR(OPEN, VOLUME, 10))} \\
\small 140 & \small Volume \& Flow & \small Min Rank of Intraday Position vs. Vol Correlation & x & \fml{MIN(RANK(DECAYLINEAR(((RANK(OPEN) + RANK(LOW)) - (RANK(HIGH) + RANK(CLOSE))), 8)), TSRANK(DECAYLINEAR(CORR(TSRANK(CLOSE, 8), TSRANK(MEAN(VOLUME,60), 20), 8), 7), 3))} \\
\small 141 & \small Volume \& Flow & \small Inverse Rank of High-Price and Volume Correlation &  & \fml{(RANK(CORR(RANK(HIGH), RANK(MEAN(VOLUME,15)), 9))* -1)} \\
\small 142 & \small Trend \& Momentum & \small Triple Rank: Price Recency, Acceleration, and Volume &  & \fml{(((-1 * RANK(TSRANK(CLOSE, 10))) * RANK(DELTA(DELTA(CLOSE, 1), 1))) * RANK(TSRANK((VOLUME /MEAN(VOLUME,20)), 5)))} \\
\small 143 & \small Trend \& Momentum & \small Conditional Return Accumulation &  & \fml{CLOSE>DELAY(CLOSE,1)?(CLOSE-DELAY(CLOSE,1))/DELAY(CLOSE,1)*SELF:SELF} \\
\small 144 & \small Liquidity \& VWAP & \small Down-Day Price Impact & x & \fml{SUMIF(ABS(CLOSE/DELAY(CLOSE,1)-1)/AMOUNT,20,CLOSE<DELAY(CLOSE,1))/COUNT(CLOSE<DELAY(CLOSE,1),20)} \\
\small 145 & \small Volume \& Flow & \small Percentage Volume Oscillator &  & \fml{(MEAN(VOLUME,9)-MEAN(VOLUME,26))/MEAN(VOLUME,12)*100} \\
\small 146 & \small Volatility \& Risk & \small Residual Return Variance vs. Mean Deviation &  & \fml{MEAN((CLOSE-DELAY(CLOSE,1))/DELAY(CLOSE,1)-SMA((CLOSE-DELAY(CLOSE,1))/DELAY(CLOSE,1),61,2),20)*((CLOSE-DELAY(CLOSE,1))/DELAY(CLOSE,1)-SMA((CLOSE-DELAY(CLOSE,1))/DELAY(CLOSE,1),61,2))/SMA(((CLOSE-DELAY(CLOSE,1))/DELAY(CLOSE,1)-((CLOSE-DELAY(CLOSE,1))/DELAY(CLOSE,1)-SMA((CLOSE-DELAY(CLOSE,1))/DELAY(CLOSE,1),61,2)))^2,60)} \\
\small 147 & \small Trend \& Momentum & \small 12-Day Regression Slope of Mean Price &  & \fml{REGBETA(MEAN(CLOSE,12),SEQUENCE(12))} \\
\small 148 & \small Volume \& Flow & \small Open Price Position vs. Volume Correlation Rank &  & \fml{((RANK(CORR((OPEN), SUM(MEAN(VOLUME,60), 9), 6)) < RANK((OPEN - TSMIN(OPEN, 14)))) * -1)} \\
\small 149 & \small Volatility \& Risk & \small Bear Market Beta &  & \fml{REGBETA(FILTER(CLOSE/DELAY(CLOSE,1)-1,BANCHMARKINDEXCLOSE<DELAY(BANCHMARKINDEXCLOSE,1)),FILTER(BANCHMARKINDEXCLOSE/DELAY(BANCHMARKINDEXCLOSE,1)-1,BANCHMARKINDEXCLOSE<DELAY(BANCHMARKINDEXCLOSE,1)),252)} \\
\small 150 & \small Volume \& Flow & \small Typical Price-Volume Flow &  & \fml{(CLOSE+HIGH+LOW)/3*VOLUME} \\
\small 151 & \small Trend \& Momentum & \small 20-Day SMA of 20-Day Momentum &  & \fml{SMA(CLOSE-DELAY(CLOSE,20),20,1)} \\
\small 152 & \small Trend \& Momentum & \small Dual SMA Crossover of Nested Price Ratios &  & \fml{SMA(MEAN(DELAY(SMA(DELAY(CLOSE/DELAY(CLOSE,9),1),9,1),1),12)-MEAN(DELAY(SMA(DELAY(CLOSE/DELAY(CLOSE,9),1),9,1),1),26),9,1)} \\
\small 153 & \small Mean Reversion & \small Multi-MA Average &  & \fml{(MEAN(CLOSE,3)+MEAN(CLOSE,6)+MEAN(CLOSE,12)+MEAN(CLOSE,24))/4} \\
\small 154 & \small Liquidity \& VWAP & \small VWAP Recovery vs. Volume Correlation Logic &  & \fml{(((VWAP - MIN(VWAP, 16))) < (CORR(VWAP, MEAN(VOLUME,180), 18)))} \\
\small 155 & \small Volume \& Flow & \small Volume MACD Histogram &  & \fml{SMA(VOLUME,13,2)-SMA(VOLUME,27,2)-SMA(SMA(VOLUME,13,2)-SMA(VOLUME,27,2),10,2)} \\
\small 156 & \small Liquidity \& VWAP & \small Inverse Rank of Decayed VWAP and Open/Low Deltas &  & \fml{(MAX(RANK(DECAYLINEAR(DELTA(VWAP, 5), 3)), RANK(DECAYLINEAR(((DELTA(((OPEN * 0.15) + (LOW *0.85)), 2) / ((OPEN * 0.15) + (LOW * 0.85))) * -1), 3))) * -1)} \\
\small 157 & \small Trend \& Momentum & \small Log-Sum Rank Product of Price Deltas &  & \fml{(MIN(PROD(RANK(RANK(LOG(SUM(TSMIN(RANK(RANK((-1 * RANK(DELTA((CLOSE - 1), 5))))), 2), 1)))), 1), 5) + TSRANK(DELAY((-1 * RET), 6), 5))} \\
\small 158 & \small Volatility \& Risk & \small Normalized Range Relative to SMA &  & \fml{((HIGH-SMA(CLOSE,15,2))-(LOW-SMA(CLOSE,15,2)))/CLOSE} \\
\small 159 & \small Mean Reversion & \small Weighted Triple-Window Oversold Oscillator &  & \fml{((CLOSE-SUM(MIN(LOW,DELAY(CLOSE,1)),6))/SUM(MAX(HGIH,DELAY(CLOSE,1))-MIN(LOW,DELAY(CLOSE,1)),6)*12*24+(CLOSE-SUM(MIN(LOW,DELAY(CLOSE,1)),12))/SUM(MAX(HGIH,DELAY(CLOSE,1))-MIN(LOW,DELAY(CLOSE,1)),12)*6*24+(CLOSE-SUM(MIN(LOW,DELAY(CLOSE,1)),24))/SUM(MAX(HGIH,DELAY(CLOSE,1))-MIN(LOW,DELAY(CLOSE,1)),24)*6*24)*100/(6*12+6*24+12*24)} \\
\small 160 & \small Volatility \& Risk & \small 20-Day SMA of Downside Volatility &  & \fml{SMA((CLOSE<=DELAY(CLOSE,1)?STD(CLOSE,20):0),20,1)} \\
\small 161 & \small Volatility \& Risk & \small 12-Day Average True Range &  & \fml{MEAN(MAX(MAX((HIGH-LOW),ABS(DELAY(CLOSE,1)-HIGH)),ABS(DELAY(CLOSE,1)-LOW)),12)} \\
\small 162 & \small Trend \& Momentum & \small 12-Day Stochastic RSI &  & \fml{(SMA(MAX(CLOSE-DELAY(CLOSE,1),0),12,1)/SMA(ABS(CLOSE-DELAY(CLOSE,1)),12,1)*100-MIN(SMA(MAX(CLOSE-DELAY(CLOSE,1),0),12,1)/SMA(ABS(CLOSE-DELAY(CLOSE,1)),12,1)*100,12))/(MAX(SMA(MAX(CLOSE-DELAY(CLOSE,1),0),12,1)/SMA(ABS(CLOSE-DELAY(CLOSE,1)),12,1)*100,12)-MIN(SMA(MAX(CLOSE-DELAY(CLOSE,1),0),12,1)/SMA(ABS(CLOSE-DELAY(CLOSE,1)),12,1)*100,12))} \\
\small 163 & \small Volume \& Flow & \small Rank of Overbought Shadow x Volume Flow &  & \fml{RANK(((((-1 * RET) * MEAN(VOLUME,20)) * VWAP) * (HIGH - CLOSE)))} \\
\small 164 & \small Volatility \& Risk & \small 13-Day SMA of Inverse Volatility Intensity &  & \fml{SMA((((CLOSE>DELAY(CLOSE,1))?1/(CLOSE-DELAY(CLOSE,1)):1)-MIN(((CLOSE>DELAY(CLOSE,1))?1/(CLOSE-DELAY(CLOSE,1)):1),12))/(HIGH-LOW)*100,13,2)} \\
\small 165 & \small Volatility \& Risk & \small 48-Day Rescaled Range & x & \fml{MAX(SUMAC(CLOSE-MEAN(CLOSE,48)))-MIN(SUMAC(CLOSE-MEAN(CLOSE,48)))/STD(CLOSE,48)} \\
\small 166 & \small Volatility \& Risk & \small 20-Day Skewness of Price Returns &  & \fml{-20*(20-1)^1.5*SUM(CLOSE/DELAY(CLOSE,1)-1-MEAN(CLOSE/DELAY(CLOSE,1)-1,20),20)/((20-1)*(20-2)(SUM((CLOSE/DELAY(CLOSE,1),20)^2,20))^1.5)} \\
\small 167 & \small Trend \& Momentum & \small 12-Day Cumulative Positive Price Change &  & \fml{SUM((CLOSE-DELAY(CLOSE,1)>0?CLOSE-DELAY(CLOSE,1):0),12)} \\
\small 168 & \small Volume \& Flow & \small Negative Relative Volume &  & \fml{(-1*VOLUME/MEAN(VOLUME,20))} \\
\small 169 & \small Trend \& Momentum & \small Dual SMA Crossover of Price Deltas &  & \fml{SMA(MEAN(DELAY(SMA(CLOSE-DELAY(CLOSE,1),9,1),1),12)-MEAN(DELAY(SMA(CLOSE-DELAY(CLOSE,1),9,1),1),26),10,1)} \\
\small 170 & \small Volume \& Flow & \small Complex Volume-Weighted Price Position Rank &  & \fml{((((RANK((1 / CLOSE)) * VOLUME) / MEAN(VOLUME,20)) * ((HIGH * RANK((HIGH - CLOSE))) / (SUM(HIGH, 5) / 5))) - RANK((VWAP - DELAY(VWAP, 5))))} \\
\small 171 & \small Mean Reversion & \small Power-Weighted Intraday Bias Ratio &  & \fml{((-1 * ((LOW - CLOSE) * (OPEN^5))) / ((CLOSE - HIGH) * (CLOSE^5)))} \\
\small 172 & \small Trend \& Momentum & \small 6-Day Average Directional Index &  & \fml{MEAN(ABS(SUM((LD>0 \& LD>HD)?LD:0,14)*100/SUM(TR,14)-SUM((HD>0 \& HD>LD)?HD:0,14)*100/SUM(TR,14))/(SUM((LD>0 \& LD>HD)?LD:0,14)*100/SUM(TR,14)+SUM((HD>0 \& HD>LD)?HD:0,14)*100/SUM(TR,14))*100,6)} \\
\small 173 & \small Trend \& Momentum & \small Triple Exponential Price Smoothing &  & \fml{3*SMA(CLOSE,13,2)-2*SMA(SMA(CLOSE,13,2),13,2)+SMA(SMA(SMA(LOG(CLOSE),13,2),13,2),13,2)} \\
\small 174 & \small Volatility \& Risk & \small 20-Day SMA of Upside Volatility &  & \fml{SMA((CLOSE>DELAY(CLOSE,1)?STD(CLOSE,20):0),20,1)} \\
\small 175 & \small Volatility \& Risk & \small 6-Day Average True Range &  & \fml{MEAN(MAX(MAX((HIGH-LOW),ABS(DELAY(CLOSE,1)-HIGH)),ABS(DELAY(CLOSE,1)-LOW)),6)} \\
\small 176 & \small Volume \& Flow & \small Correlation of Stochastic Position and Volume &  & \fml{CORR(RANK(((CLOSE - TSMIN(LOW, 12)) / (TSMAX(HIGH, 12) - TSMIN(LOW,12)))), RANK(VOLUME), 6)} \\
\small 177 & \small Trend \& Momentum & \small 20-Day High-Day Recency &  & \fml{((20-HIGHDAY(HIGH,20))/20)*100} \\
\small 178 & \small Volume \& Flow & \small Daily Volume-Weighted Return &  & \fml{(CLOSE-DELAY(CLOSE,1))/DELAY(CLOSE,1)*VOLUME} \\
\small 179 & \small Volume \& Flow & \small Rank of VWAP-Volume and Low-Volume Corrs &  & \fml{(RANK(CORR(VWAP, VOLUME, 4)) *RANK(CORR(RANK(LOW), RANK(MEAN(VOLUME,50)), 12)))} \\
\small 180 & \small Trend \& Momentum & \small Volume-Conditional Momentum Rank &  & \fml{((MEAN(VOLUME,20) < VOLUME) ? ((-1 * TSRANK(ABS(DELTA(CLOSE, 7)), 60)) * SIGN(DELTA(CLOSE, 7)) : (-1 * VOLUME)))} \\
\small 181 & \small Volatility \& Risk & \small Benchmark-Relative Excess Return Skewness &  & \fml{SUM(((CLOSE/DELAY(CLOSE,1)-1)-MEAN((CLOSE/DELAY(CLOSE,1)-1),20))-(BANCHMARKINDEXCLOSE-MEAN(BANCHMARKINDEXCLOSE,20))^2,20)/SUM((BANCHMARKINDEXCLOSE-MEAN(BANCHMARKINDEXCLOSE,20))^3)} \\
\small 182 & \small Volatility \& Risk & \small Market Synchronization Frequency &  & \fml{COUNT((CLOSE>OPEN \& BANCHMARKINDEXCLOSE>BANCHMARKINDEXOPEN)OR(CLOSE<OPEN \& BANCHMARKINDEXCLOSE<BANCHMARKINDEXOPEN),20)/20} \\
\small 183 & \small Volatility \& Risk & \small 24-Day Rescaled Range & x & \fml{MAX(SUMAC(CLOSE-MEAN(CLOSE,24)))-MIN(SUMAC(CLOSE-MEAN(CLOSE,24)))/STD(CLOSE,24)} \\
\small 184 & \small Trend \& Momentum & \small Rank of Delayed Price-Gap Correlation &  & \fml{(RANK(CORR(DELAY((OPEN - CLOSE), 1), CLOSE, 200)) + RANK((OPEN - CLOSE)))} \\
\small 185 & \small Volatility \& Risk & \small Rank of Squared Intraday Return &  & \fml{RANK((-1 * ((1 - (OPEN / CLOSE))^2)))} \\
\small 186 & \small Trend \& Momentum & \small 12-Day Smoothed Average Directional Index &  & \fml{(MEAN(ABS(SUM((LD>0 \& LD>HD)?LD:0,14)*100/SUM(TR,14)-SUM((HD>0 \& HD>LD)?HD:0,14)*100/SUM(TR,14))/(SUM((LD>0 \& LD>HD)?LD:0,14)*100/SUM(TR,14)+SUM((HD>0 \& HD>LD)?HD:0,14)*100/SUM(TR,14))*100,6)+DELAY(MEAN(ABS(SUM((LD>0 \& LD>HD)?LD:0,14)*100/SUM(TR,14)-SUM((HD>0 \& HD>LD)?HD:0,14)*100/SUM(TR,14))/(SUM((LD>0 \& LD>HD)?LD:0,14)*100/SUM(TR,14)+SUM((HD>0 \& HD>LD)?HD:0,14)*100/SUM(TR,14))*100,6),6))/2} \\
\small 187 & \small Volatility \& Risk & \small 20-Day Cumulative Upward Gap Pressure &  & \fml{SUM((OPEN<=DELAY(OPEN,1)?0:MAX((HIGH-OPEN),(OPEN-DELAY(OPEN,1)))),20)} \\
\small 188 & \small Volatility \& Risk & \small 11-Day Relative Range Deviation &  & \fml{((HIGH-LOW-SMA(HIGH-LOW,11,2))/SMA(HIGH-LOW,11,2))*100} \\
\small 189 & \small Volatility \& Risk & \small 6-Day Mean Absolute Deviation &  & \fml{MEAN(ABS(CLOSE-MEAN(CLOSE,6)),6)} \\
\small 190 & \small Volatility \& Risk & \small Log Gain-to-Loss Variability Ratio &  & \fml{LOG((COUNT(CLOSE/DELAY(CLOSE)-1>((CLOSE/DELAY(CLOSE,19))^(1/20)-1),20)-1)*(SUMIF(((CLOSE/DELAY(CLOSE)-1-(CLOSE/DELAY(CLOSE,19))^(1/20)-1))^2,20,CLOSE/DELAY(CLOSE)-1<(CLOSE/DELAY(CLOSE,19))^(1/20)-1))/((COUNT((CLOSE/DELAY(CLOSE)-1<(CLOSE/DELAY(CLOSE,19))^(1/20)-1),20))*(SUMIF((CLOSE/DELAY(CLOSE)-1-((CLOSE/DELAY(CLOSE,19))^(1/20)-1))^2,20,CLOSE/DELAY(CLOSE)-1>(CLOSE/DELAY(CLOSE,19))^(1/20)-1))))} \\
\small 191 & \small Mean Reversion & \small Volume-Low Corr Adjusted Mid-Price Spread &  & \fml{((CORR(MEAN(VOLUME,20), LOW, 5) + ((HIGH + LOW) / 2)) - CLOSE)} \\
\end{longtable}
\end{landscape}

\end{document}